\documentclass[12pt]{article} \usepackage[margin=1in]{geometry} \usepackage[authoryear,round,comma]{natbib}

\usepackage{graphicx}
\usepackage{multirow}
\usepackage{amsmath,amssymb,amsfonts}
\usepackage{amsthm}
\usepackage{mathrsfs}              
\usepackage[title]{appendix}       
\usepackage{xcolor}
\usepackage{textcomp}
\usepackage{manyfoot}               
\usepackage{booktabs}              
\usepackage{array}                
\usepackage{url}
\providecommand{\doi}[1]{\url{https://doi.org/#1}}

\graphicspath{{figures/}{figures/ladder_circuits/}}

\theoremstyle{plain}

\theoremstyle{remark}

\theoremstyle{definition}

\begin{document}

\title{OQMD: Single-Qubit Rotation Control Improves
  Low-CNOT Multiclass Quantum Classification}

\author{
  Michael A.\ Magid$^{1,*}$,\quad
  Melissa Zeynep Ertem$^{1}$,\quad
  Jun Suzuki$^{2,3}$\\[0.6em]
  \parbox{0.88\textwidth}{\footnotesize\centering
  $^{1}$School of Systems Science and Industrial Engineering,
  Thomas J.\ Watson College of Engineering and Applied Science,
  Binghamton University, State University of New York,
  Binghamton, NY 13902-6000, USA\\[0.25em]
  $^{2}$Graduate School of Informatics and Engineering\quad
  $^{3}$Institute for Advanced Science,
  The University of Electro-Communications,
  1-5-1 Chofugaoka, Chofu-shi, Tokyo 182-8585, Japan\\[0.25em]
  $^{*}$Corresponding author:\ \texttt{mmagid1@binghamton.edu}%
  }%
}
\date{}
\maketitle

\begin{abstract}
Near-term variational classifiers incur substantial error and latency
from two-qubit gates, yet practitioners often assume that additional
entangling depth is the default route to higher accuracy.
This work studies Optimal Quantum Measurement Decoding (OQMD):
optimizing how quantum outcomes are mapped to classical labels by
training a readout layer before measurement, jointly with the
variational circuit, without adding CNOTs.
Experiments use trainable triple single-qubit rotations as one
concrete, hardware-native realization of OQMD; other single-qubit
parametrizations fit the same classical outer loop.
On the Iris benchmark with a 30-point stratified test split, the best
observed 0-CNOT configuration with OQMD reaches 83.33\% accuracy, with a 96\% at 9 CNOTs, exceeding the best 18-CNOT controls (56.67\%) and the best 18-CNOT
configuration with OQMD (66.67\%) under a common protocol.
A six-point CNOT-depth series from 0 to 18 (fixed optimizer, iteration
budget, random-seed count, and ZXZ readout) shows that the highest raw
scores need not occur at the largest template, so aggregate complexity
is not summarized by CNOT count alone.
Because run-level accuracies are discrete and non-Gaussian, we
emphasize best-observed scores and, where a global comparison of
pooled runs is required, Mann--Whitney $U$ tests rather than parametric
tests on means.
Across architectures, OQMD shows statistically consistent but
magnitude-dependent gains: large peak lifts on minimal circuits coexist
with a small pooled mean shift on complex 18-CNOT runs
($p\approx 0.03$) that is not ``universal'' in the sense of uniformly
large practical effects.%
\end{abstract}

\maketitle

\section{Introduction}
Variational quantum classifiers (VQCs) combine parameterized quantum circuits with a classical optimizer for hybrid training~\cite{cerezo2021variational, bharti2022noisy}. They use superposition and entanglement and may offer advantages on some learning problems relative to classical baselines~\cite{biamonte2017quantum, dunjko2018machine}, but realized performance depends strongly on circuit design and hyperparameters.

Entanglement is widely treated as a central resource in quantum machine learning; empirical work shows settings where entangled representations can reduce prediction error and improve performance~\cite{wang2024transition}. In variational circuits, entanglement is typically introduced through multi-qubit gates such as CNOTs, which motivates the common design practice of increasing entangling depth to improve accuracy.

A fundamental challenge facing quantum machine learning in the Noisy Intermediate-Scale Quantum (NISQ) era is the severe limitation imposed by two-qubit entangling gates, particularly CNOT gates. Current quantum hardware exhibits CNOT error rates 10-100 times higher than single-qubit gates, with execution times 10-50 times longer, leading to increased decoherence and error accumulation~\cite{preskill2018quantum}. These constraints create a critical tension: while circuit expressivity typically requires entanglement through CNOT gates, practical deployability demands minimal CNOT usage to reduce errors. This motivates the development of techniques that improve classification accuracy without adding CNOT gates.

We introduce Optimal Quantum Measurement Decoding (OQMD), a measurement-decoding framework: the classical optimizer learns parameters of a readout map from measured qubit outcomes to class labels, jointly with the variational circuit, so that the mapping from quantum data to classical predictions is trained rather than fixed by default parity rules alone.
Here we instantiate that readout with trainable single-qubit rotations immediately before measurement (adding no CNOTs); this is not the only possible realization---for example, standard $\mathrm{SU}(2)$ product gates could be substituted, and full measurement optimization may ultimately involve collective measurements (left to future work).
The size of the improvement from OQMD still depends on architecture (pooled shifts are statistically consistent under Mann--Whitney $U$, but mean gaps need not be large; Section~\ref{sec:results}).

The performance of variational quantum classifiers is highly sensitive to architectural choices, including feature map selection, ansatz design, rotation gate strategies, and optimization algorithms~\cite{havlicek2019supervised, schuld2020circuit}. Among these design choices, rotation gate sequences influence circuit expressivity and trainability by determining accessible regions of Hilbert space and the resulting optimization landscape~\cite{du2020expressive, holmes2022connecting, sim2019expressibility}. Much of the literature still relies on heuristic templates or small empirical comparisons rather than full factorials over readout and ansatz choices~\cite{larose2019variational, cerezo2021cost}.

In this paper, all other design choices (feature map, ansatz, optimizer, and classical outer loop) are fixed to explicit baselines so that reported gains isolate the readout layer; we use the triple-rotation realization as a concrete, hardware-friendly implementation of OQMD rather than as the definition of OQMD itself.
The Iris dataset provides a small, controlled setting in which differences in peak hold-out accuracy primarily reflect circuit layout, optimization, and measurement decoding rather than dataset shift or large-scale data effects.
Our results challenge the informal expectation that ``more CNOTs always help'': the largest raw scores arise on low- and mid-depth circuit templates when OQMD is enabled, while pooled CONTROL versus OQMD comparisons remain favorable under Mann--Whitney $U$ on run-level accuracies (Section~\ref{sec:results}).

The paper makes four main contributions: (i)~peak-focused benchmarks across minimal (0 CNOT), intermediate (structured mid-depth entanglement), and complex (18 CNOT) families; (ii)~a six-point CNOT-depth series in which non-readout hyperparameters are held fixed (COBYLA, 200 iterations, 50 random seeds per condition, ZXZ OQMD) so that accuracy can be compared to hardware cost; (iii)~factorial sweeps over readout parameterizations (triple rotations) on minimal and complex designs, showing that sensitivity to that choice depends on architecture; and (iv)~analysis and plots appropriate for non-Gaussian run distributions, with modes noted in Section~\ref{sec:results} rather than Gaussian means in headlines.

Classical reference accuracies for Iris are summarized in Appendix~\ref{app:classical_comparison} when interpreting the magnitude of the quantum scores.

\section{Related work}

Variational quantum classifiers (VQCs) employ hybrid quantum-classical optimization of parameterized circuits~\cite{cerezo2021variational, schuld2020circuit}. While early demonstrations showed promise~\cite{havlicek2019supervised}, systematic architectural studies remain limited. The expressivity-trainability trade-off is fundamental: highly expressive circuits risk barren plateaus~\cite{mcclean2018barren}, while simple circuits may underfit~\cite{du2020expressive, sim2019expressibility}. Systematic categorization of QML methods across NISQ and fault-tolerant regimes remains an active area~\cite{majid2025systematic}. Measurement optimization can improve accuracy at fixed CNOT budget without increasing two-qubit depth.

NISQ-era constraints necessitate gate-efficient designs. Hardware-efficient ansätze balance expressivity with native gate sets~\cite{kandala2017hardware}, while compilation techniques reduce CNOT count~\cite{nam2018automated}. Recent VQC applications demonstrate that feature map and ansatz choices have strong practical impact on classification accuracy~\cite{regadio2025exoplanet}. However, these approaches focus on state preparation. OQMD complements them by optimizing the measurement basis itself---an underexplored degree of freedom related to decoding and readout design~\cite{mahler2013adaptive, higgins2007entanglement}. A complementary direction improves accuracy by aggregating multiple VQCs through learned weighting rather than modifying the readout layer of a single circuit~\cite{tolotti2025weighted}. Our experiments focus on empirical optimization performance rather than information-geometric characterizations of the parameter space.

\section{Methodology}
\label{sec:methodology}

\subsection{Dataset and preprocessing}

We use the Iris dataset from scikit-learn~\cite{pedregosa2011scikit}: 150 samples of three species (\textit{I. setosa}, \textit{I. versicolor}, and \textit{I. virginica}), each with four features (sepal length and width, petal length and width, in centimeters). A single stratified split yields 120 training and 30 test samples (80\%/20\%). Fixing this split and preprocessing isolates changes in the quantum pipeline from data curation effects.

Features are min--max scaled to $[0, 1]$ for angle encoding. All four features are retained (no dimensionality reduction).

\subsection{VQC architectures}

We use three representative VQC designs ordered by structural complexity rather than by CNOT count alone: a minimal design (no CNOT gates), an intermediate design (four configurations at 3, 6, 9, and 12 CNOT), and a complex design (CNOTs for both feature map and ansatz). 

\subsubsection{Architecture 1: Minimal circuits (0 CNOT)}

This design evaluates OQMD under severe resource constraints, as in shallow or connectivity-limited devices.
The feature map is chosen as one of the simplest ones,
$$U_{\text{ZFM}}(\mathbf{x}) = \prod_{i=1}^{4} R_Z(x_i) \cdot H^{\otimes 4}$$
This is called the ZFeatureMap (ZFM), which is equivalent to the Qiskit phase gate $P$ up to the convention $P(\theta)=e^{i\theta/2}R_Z(\theta)$.
The ansatz is set to a product of single rotations around $y$-axis (SingleLayerRY, SLRY).
To simplify the notation, we shall identify this VQC by `ZFM(1)/SLRY(1)'.
It should be emphasized that this VQC does not utilize \textit{any} entanglement.


\subsubsection{Architecture 2: Intermediate circuits (structured mid-depth)}

Architecture~2 is not a single circuit but a \textit{family} of four ladder templates between the 0-CNOT minimal design and the 18-CNOT complex design (Table~\ref{tab:six_cnot_configs}). Three rungs (3, 6, and 9 CNOT) keep the same ZFeatureMap encoding, $U_{\text{ZFM}}(\mathbf{x}) = \prod_i R_Z(x_i)\,H^{\otimes 4}$, and increase only RealAmplitudes depth: ZFM(1)/RA(1), ZFM(1)/RA(2), and ZFM(1)/RA(3).
The last rung implements the ZZ feature map (ZZFM),
$$U_{\text{ZZ}}(\mathbf{x}) = \prod_{k=1}^{2} \left[ \prod_{i=1}^{4} R_Z(x_i) \cdot \prod_{\langle i, j \rangle} \text{CZ}_{i,j} \cdot R_{ZZ}(x_i x_j) \right] H^{\otimes 4}$$
which entangles two qubits conditioned on input data.
This 12-CNOT rung switches to ZZFeatureMap with two repetitions and a SingleLayerRY ansatz, ZZFM(2)/SLRY(1); this rung changes \textit{both} the feature map and the ansatz relative to the 3--9 CNOT templates, so it is not a pure depth extension of the same encoding--ansatz pair.
\begin{itemize}
    \item ZFM(1)/RA(1) [3 CNOT]
    \item ZFM(1)/RA(2) [6 CNOT]
    \item ZFM(1)/RA(3) [9 CNOT]
    \item ZZFM(2)/SLRY(1) [12 CNOT]
\end{itemize}
Trainable ansatz parameters on four qubits with linear entanglement are 8, 12, and 16 for RA(1--3), and four for SLRY at the 12-CNOT rung (feature-map parameters remain data-dependent).


\subsubsection{Architecture 3: Complex circuits (18 CNOT)}

This design evaluates OQMD on a heavily entangled template typical of variational classifiers in the literature.
To have an efficient encoding, the ZZFMap is repeated twice.
As an ansatz, the RA is also repeated twice. The total number of CNOT gates is 18 in total. 



Table~\ref{tab:experimental_design} summarizes the three architectural families (feature map, ansatz, CNOT budget, trainable parameters, and factorial scope). The four intermediate ladder templates (3, 6, 9, and 12 CNOT) are defined in Section~\ref{sec:methodology} and drawn gate-by-gate in Appendix~\ref{app:circuit_definitions}.

Figure~\ref{fig:circuit_architectures} shows the complete circuit diagrams for architectures~1 and~3, illustrating the gate-level implementation and CNOT placement.

\begin{figure*}[htbp]
 \centering
 \includegraphics[width=0.48\textwidth]{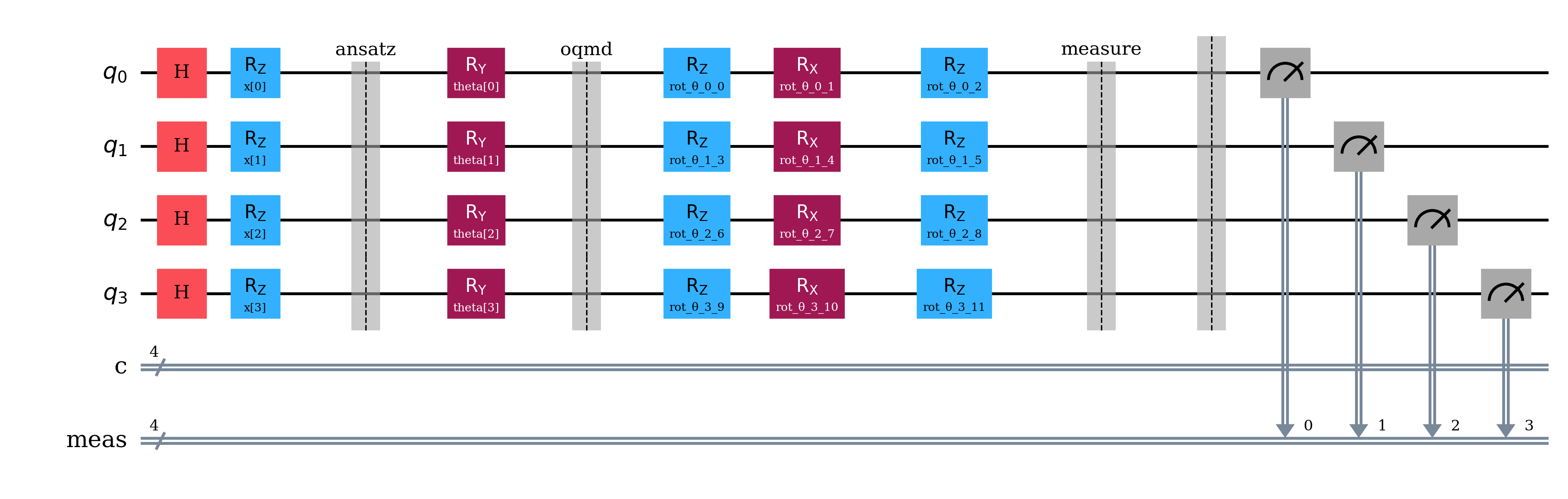}
 \hfill
 \includegraphics[width=0.48\textwidth]{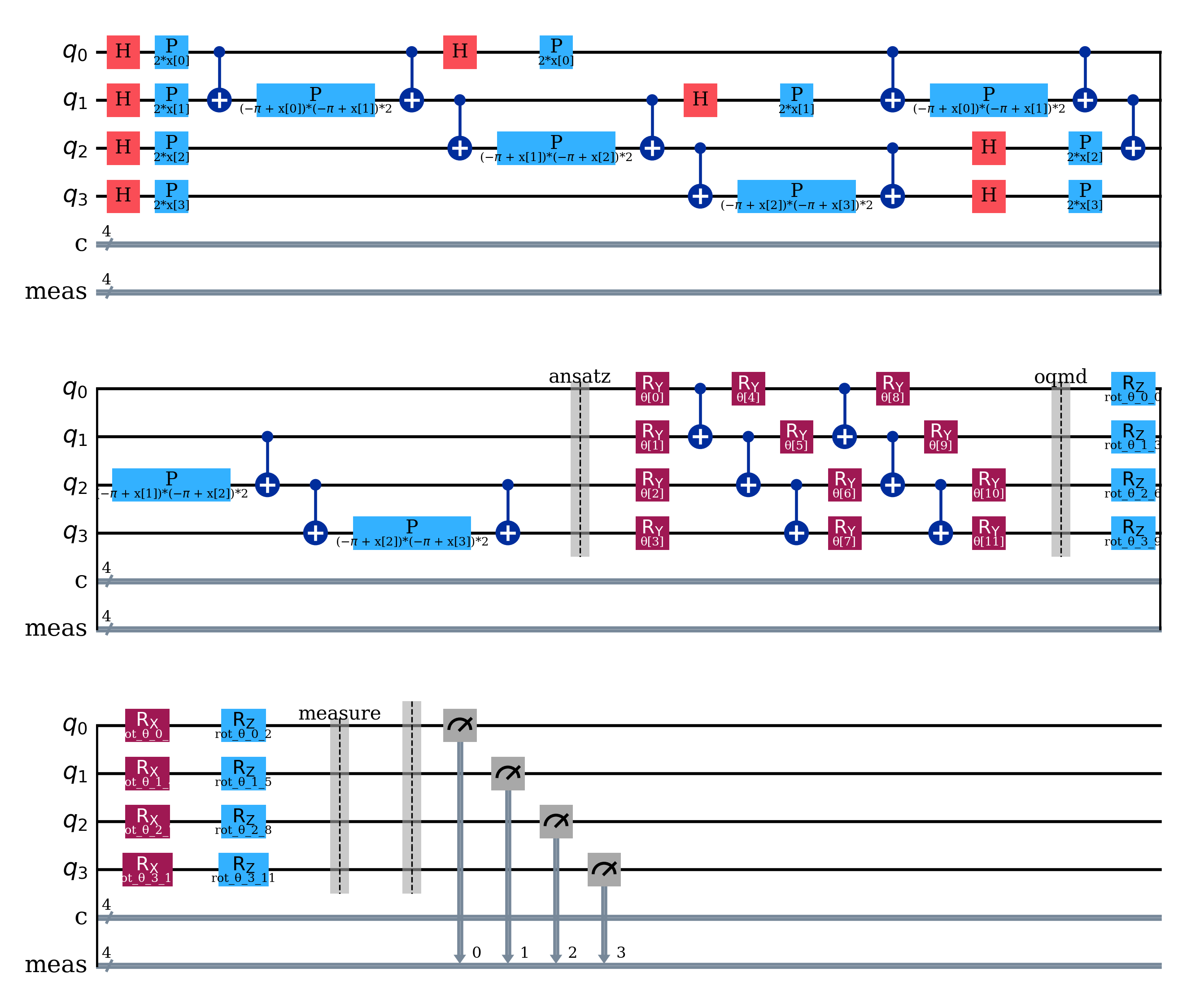}
 \caption{Circuit diagrams for Architecture~1 (minimal): $H$ + $R_Z$ feature encoding, single $R_Y$ ansatz layer, ZXZ OQMD rotations, 0 CNOT gates (left). Architecture~3 (complex): ZZFeatureMap (12 CNOT) + RealAmplitudes (6 CNOT) + OQMD (0 CNOT) = 18 CNOT total (right). OQMD adds measurement optimization with zero CNOT overhead in both cases.}
 \label{fig:circuit_architectures}
\end{figure*}

\subsection{OQMD: Optimal Quantum Measurement Decoding}

\subsubsection{Theoretical foundation}

Standard VQC measurement projects the quantum state onto the computational basis $\{|0\rangle, |1\rangle\}^{\otimes n}$. OQMD generalizes this decoding scheme by choosing measurement axes for each qubit.
This is easily implemented by first applying parameterized rotation gates followed by the standard $Z$ measurements.
Let $U_{\text{OQMD}}(\boldsymbol{\phi})$ be a product of rotation gates for each qubit with trainable angles $\boldsymbol{\phi}$.
The total VQC with OQMD is given as
$$|\psi_{\text{measured}}\rangle = U_{\text{OQMD}}(\boldsymbol{\phi}) \cdot U_{\text{ansatz}}(\boldsymbol{\theta}) \cdot U_{\text{feature}}(\mathbf{x}) |0\rangle^{\otimes n}$$
where $U_{\text{feature}}(\mathbf{x})$ encodes input data and $U_{\text{ansatz}}(\boldsymbol{\theta})$ is the variational ansatz with trainable parameters $\boldsymbol{\theta}$.

The core of OQMD is that the measurement angles $\boldsymbol{\phi}$ can be treated as trainable parameters.
Hence, the complete parameter vector $\boldsymbol{\Theta} = (\boldsymbol{\theta}, \boldsymbol{\phi})$ is jointly optimized to minimize the classification loss. Pre-measurement rotation strategies of this kind represent an emerging design pattern in quantum classification~\cite{sahoo2026entanglement}.

Figure~\ref{fig:oqmd_concept} illustrates the general OQMD concept: standard VQC measures in the computational basis, while OQMD first rotates into an optimized measurement basis, effectively learning both state preparation and optimal observables. Figure~\ref{fig:control_vs_oqmd_circuits} contrasts CONTROL vs.\ OQMD on Architecture~1 (minimal) as a concrete example.
\begin{figure}[htbp]
 \centering
 \fbox{\parbox{0.95\columnwidth}{
 \textbf{(a) Standard VQC (CONTROL):}\\
 $|0\rangle$ -- Feature Map -- Ansatz -- $\boxed{M}$ \\[0.5em]
 \textbf{(b) VQC with OQMD:}\\
 $|0\rangle$ -- Feature Map -- Ansatz -- \colorbox{gray!20}{OQMD Rotations} -- $\boxed{M}$\\[0.5em]
 \small{OQMD adds 3 trainable rotation gates per qubit before measurement, optimizing the measurement basis with zero CNOT overhead.}
 }}
 \caption{OQMD concept: Standard VQC (a) measures in the computational basis, while OQMD (b) first applies trainable rotation gates to measure in an optimized basis. This adds expressivity without CNOT gates.}
 \label{fig:oqmd_concept}
\end{figure}
\begin{figure}[htbp]
 \centering
 \includegraphics[width=\columnwidth]{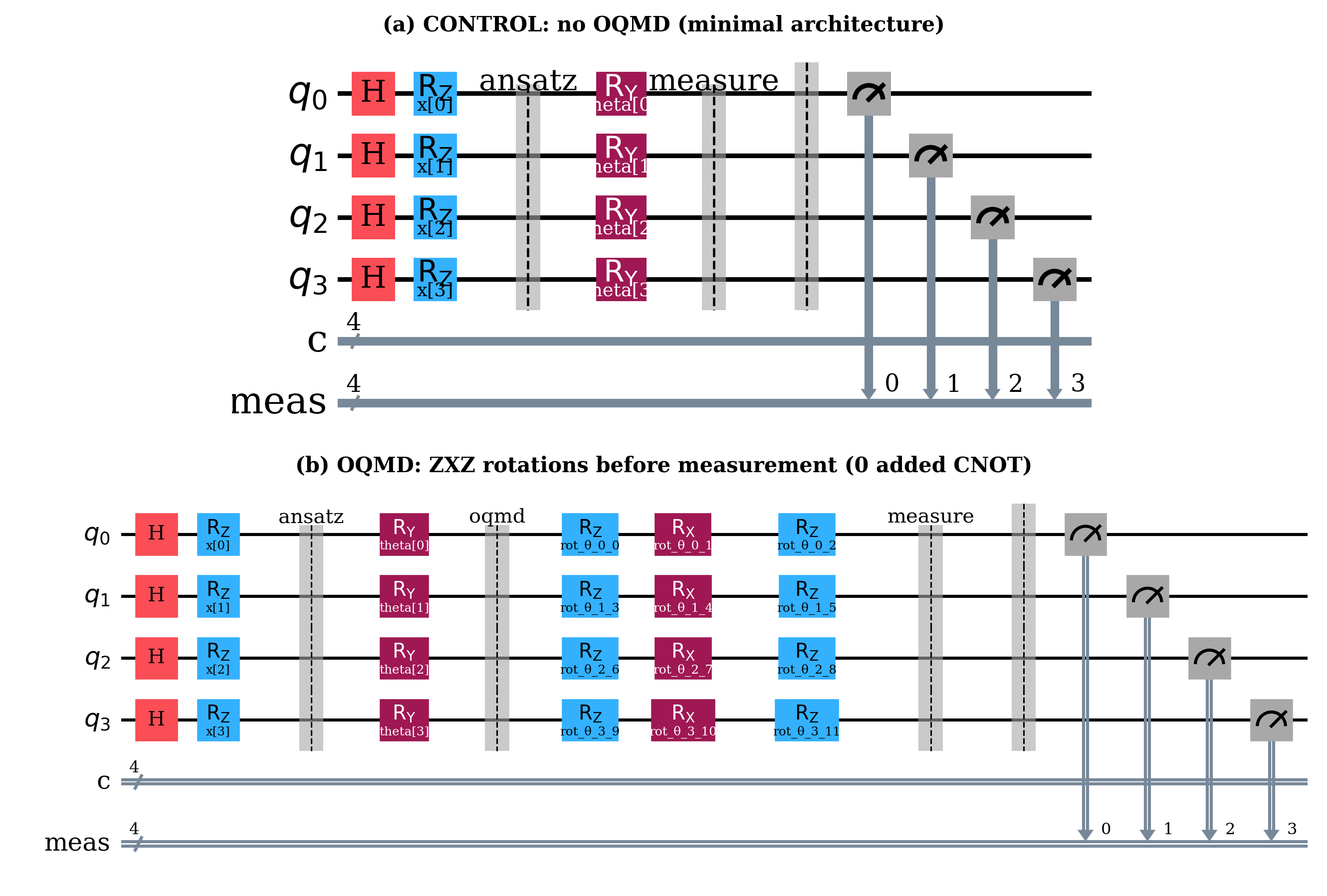}
 \caption{Detailed circuit comparison: CONTROL versus OQMD on the minimal architecture. (a) CONTROL (parity readout, no OQMD). (b) OQMD adds ZXZ triple rotations before measurement (no additional CNOTs). Peak accuracies appear in Table~\ref{tab:architecture_summary}.}
 \label{fig:control_vs_oqmd_circuits}
\end{figure}

The advantage of the proposed OQMD is immediate: For an $n$-qubit system, OQMD adds $3n$ rotation gates and $3n$ trainable parameters. All of these gates are single-qubit rotations; no CNOT gates are added. 

\subsubsection{Rotation code specification}
Upon implementing OQMD, a specific qubit rotation needs to be specified, since there is no unique way to implement it. 
Below, we consider several rotation implementation, which is called rotation codes. 
The purpose of analyzing rotation codes is to make sure whether or not OQMD depends on a specific choice of rotation codes. 
As main rotation codes, we evaluate 5 rotation strategies (4 OQMD codes + 1 control baseline), each defined by a three-character code specifying the sequence of Pauli rotations applied to each qubit before measurement:
\begin{itemize}
    \item \textbf{ZXZ}: $R_Z(\phi_1) \cdot R_X(\phi_2) \cdot R_Z(\phi_3)$ — Standard triple single-qubit gate
 \item \textbf{ZXY}: $R_Z(\phi_1) \cdot R_X(\phi_2) \cdot R_Y(\phi_3)$ — Three-axis rotation
 \item \textbf{YXY}: $R_Y(\phi_1) \cdot R_X(\phi_2) \cdot R_Y(\phi_3)$ — Y-axis dominant
 \item \textbf{YXZ}: $R_Y(\phi_1) \cdot R_X(\phi_2) \cdot R_Z(\phi_3)$ — Mixed axes
 \item \textbf{CONTROL}: No rotation gates (baseline)
\end{itemize}

In addition, eight additional rotation codes (beyond the four used on the minimal design) will be analyzed in order to gain more insight into the effect of OQMD on the complex architecture (Architecture 3). They are specified by three axes as $ZXZ\ \leftrightarrow\ R_Z(\phi_1) \cdot R_X(\phi_2) \cdot R_Z(\phi_3)$: 
\[
XYX, XYZ, XZX, XZY, YZX, YZY, ZYX, ZYZ 
\]
Thus, 18 rotation codes will be examined in total. 

%

\subsection{Quantum classifier configuration}

The VQC is configured for three-class classification with a statevector simulator (noise-free evaluation), cross-entropy loss, three output classes, and Qiskit's parity encoding from measurement outcomes to class probabilities.

The setup yields valid nonnegative class probabilities that sum to one for the three Iris species (multiclass, not one-versus-rest reduction).

\subsection{Optimization and hyperparameter search}
\label{sec:optim_search}

\subsubsection{Optimizers}

We evaluated four optimization algorithms: one gradient-free and three gradient-based methods using finite-difference gradient estimation, selected for efficiency and convergence properties:

\begin{enumerate}
 \item \textbf{COBYLA} (Constrained Optimization BY Linear Approximations): Gradient-free trust-region method for nonlinear optimization~\cite{powell1994direct}.
 
 \item \textbf{L-BFGS-B} (Limited-memory Broyden-Fletcher-Goldfarb-Shanno): Gradient-based quasi-Newton method approximating the Hessian for improved convergence~\cite{byrd1995limited}.
 
 \item \textbf{CG} (Conjugate Gradient): Gradient-based iterative method exploiting conjugate directions~\cite{hestenes1952methods}.
 
 \item \textbf{SLSQP} (Sequential Least Squares Programming): Gradient-based method for constrained optimization~\cite{kraft1988software}.
\end{enumerate}

For gradient-based methods, parameter derivatives use Qiskit's built-in circuit-gradient interface (parameter-shift style evaluations for the models here), rather than optimizing the loss as a black box.

Each optimizer was limited to a maximum of 200-205 iterations to balance computational cost and convergence quality. We excluded slower optimizers (SPSA, POWELL, TNC) that exhibited iteration limit violations during preliminary experiments, increasing computational cost by up to 4.5$\times$ without commensurate performance gains, and ADAM on grounds of per-iteration wall time on the minimal design (Appendix~\ref{app:adam_pilot}, Table~\ref{tab:adam_minimal_pilot}).

We include several classical optimizers in the factorial grids because VQC loss landscapes are not guaranteed to be well-conditioned: different algorithms emphasize gradient noise tolerance (COBYLA), quasi-Newton curvature (L-BFGS-B), cheap conjugate-direction steps (CG), or small-scale constrained structure (SLSQP). Reporting the same readout and ansatz across these choices documents whether OQMD gains depend on a single optimizer rather than on measurement decoding itself.

\subsubsection{Experimental design}

Table~\ref{tab:experimental_design} summarizes the three architectural families and how they connect to the factorial experiments vs.\ the standardized CNOT ladder (Fig.~\ref{fig:expressivity_benefit}). 
\begin{table}[t]
\centering
\footnotesize
\setlength{\tabcolsep}{3pt}
\caption{Architectural families, templates, and run counts (same Iris split). Intermediate column lists the four ladder templates at 3/6/9/12 CNOT (Section~\ref{sec:methodology}); factorial rows are rotation$\times$optimizer grids except intermediate, where only COBYLA+ZXZ ladder runs were executed. Abbreviations: ZFM = ZFeatureMap; ZZFM = ZZFeatureMap; SLRY = SingleLayerRY; RA($r$) = RealAmplitudes with $r$ repetitions (linear entanglement).}
\label{tab:experimental_design}
\begin{tabular}{@{}>{\raggedright\arraybackslash}p{0.24\columnwidth} >{\raggedright\arraybackslash}p{0.24\columnwidth} >{\raggedright\arraybackslash}p{0.24\columnwidth} >{\raggedright\arraybackslash}p{0.22\columnwidth}@{}}
\toprule
Parameter & Minimal & Interm. & Complex \\
\midrule
Feature map & ZFM(1) & ZFM(1); ZZFM(2) at 12 CNOT & ZZFM (reps=2) \\
Ansatz & SLRY(1) & RA(1--3); SLRY at 12 CNOT & RA (reps=2) \\
CNOTs & 0 & 3, 6, 9, 12 & 18 \\
Train.\ params & 4 & 8, 12, 16, 4$^{\dagger}$ & 16 \\
\midrule
Rotation factorial & 4 OQMD codes + CTRL & 1 OQMD code + CTRL & 12 OQMD codes + CTRL \\
Optimizers (fact.) & 4 & COBYLA (ladder) & 4 \\
Seeds per combo & 50 & 50 & 50 \\
\midrule
Factorial runs & 1{,}000 & 400 (ladder) & 2{,}600 \\
Ladder (50$\times$2) & 0 CNOT & 3--12 CNOT & 18 CNOT \\
\bottomrule
\multicolumn{4}{@{}p{\columnwidth}@{}}{\footnotesize $^{\dagger}$Four ansatz parameters on the 12-CNOT rung (ZZFM(2)/SLRY); feature-map parameters are data-dependent.} \\
\end{tabular}
\end{table}

Fifty random seeds per combination provide enough restarts to observe rare high-accuracy initializations on the 30-point test set while keeping the complex factorial tractable.
The iteration cap of 200 reflects a compromise between convergence and cost: pilot runs showed diminishing returns beyond roughly 150--200 COBYLA steps on these designs, and a common cap keeps learning dynamics comparable across conditions.
For the ladder, we use COBYLA because gradient-free optimization tolerates noisy objectives and uneven landscapes; after pilots it also gave stable behavior on the intermediate-depth templates, and holding the optimizer fixed isolates CNOT-related architecture from optimizer variation (the factorial studies elsewhere still compare four optimizers on minimal and complex designs).
On the ladder, OQMD is fixed to ZXZ at every CNOT level, so we do not claim a rotation factorial at intermediate depth in this paper.
The combined factorial pool comprises 3{,}600 runs (minimal and complex) for seed-sensitivity analyses; ladder statistics are reported separately.

\subsection{Training and evaluation protocol}
\label{sec:training_eval}

\subsubsection{Callback mechanism}

To track learning dynamics, we use a callback that records training progress at regular intervals:

\begin{itemize}
 \item Current parameter values $\boldsymbol{\Theta}^{(t)}$ at iteration $t$
 \item Training loss $\mathcal{L}(\boldsymbol{\Theta}^{(t)})$
 \item Hold-out test accuracy $\text{Acc}_{\text{test}}(\boldsymbol{\Theta}^{(t)})$ on a coarse schedule (not every optimizer step; see below)
\end{itemize}

The classical optimizer updates parameters using only the training objective (cross-entropy on the 120-sample training split). Callback evaluations of test accuracy are not supplied to the optimizer: they are for monitoring and visualization only. The hold-out set is not used for early stopping, hyperparameter tuning, or gradient computation.

Evaluating full hold-out accuracy requires a statevector forward pass over all 30 test points at the current parameters, so recording it at every optimizer step would dominate wall-clock time on the factorial workloads (each extra evaluation repeats the full circuit evaluation for every test sample). Our experiments therefore log hold-out accuracy on a fixed stride of every 50 optimizer iterations (plus the initial and final iterates where available), not every step---a pragmatic trade-off between curve resolution and total compute. With a 200-iteration budget, that yields at most on the order of five interior diagnostic points per run in addition to endpoints; finer resolution would require shrinking the stride, accepting longer runs, or plotting training loss every step instead. Learning curves in the appendix reflect this coarse schedule.

Callbacks use Qiskit's VQC training hooks so optimizers stay unchanged while test accuracy is logged on the fixed schedule.

\subsubsection{Evaluation metrics}
\label{sec:eval_metrics}

Primary performance metrics are as follows:

\begin{itemize}
 \item Test accuracy: fraction of correctly classified samples in the held-out test set
 \item Learning improvement: $\Delta = \text{Acc}_{\max} - \text{Acc}_{\text{start}}$
 \item Multiclass success rate: proportion of runs predicting all three classes
 \item Overfitting flag: $\text{Acc}_{\max} > \text{Acc}_{\text{final}} + 5\%$
\end{itemize}

\subsubsection{Normality check and reported summaries}
\label{sec:normality}

Run-level test accuracies on Iris (30-point split, multiples of $1/30$) are inherently discrete. On the minimal-circuit factorial pools, Shapiro--Wilk tests reject Gaussianity for both CONTROL and OQMD marginals ($p \ll 10^{-6}$). Main text figures that show mean accuracy across seeds plot 95\% confidence intervals (CI) for the mean of run-level test accuracies (Student $t$ intervals using the sample standard deviation (SD) and run count per condition), and Welch-style intervals for the difference of two independent means on mean-gain panels, not raw SD bars. Where a line summarizes mean OQMD gain versus CNOT count in a panel, it is an ordinary least squares (OLS) fit for visualization only. Tabulated $\pm$ values in bullet lists remain the sample SD for context. Where we report return on investment (ROI) in gates or parameters, it is peak accuracy gain in percentage points per added gate or parameter. Headline claims still rest on peaks, medians, and Mann--Whitney $U$ tests for global CONTROL versus OQMD contrasts; Appendix~\ref{app:mann_whitney} states the null hypothesis and role of these tests.

\section{Results}
\label{sec:results}

\subsection{Overview: factorial results and CNOT ladder}

\begin{figure}[htbp]
 \centering
 \includegraphics[width=\columnwidth]{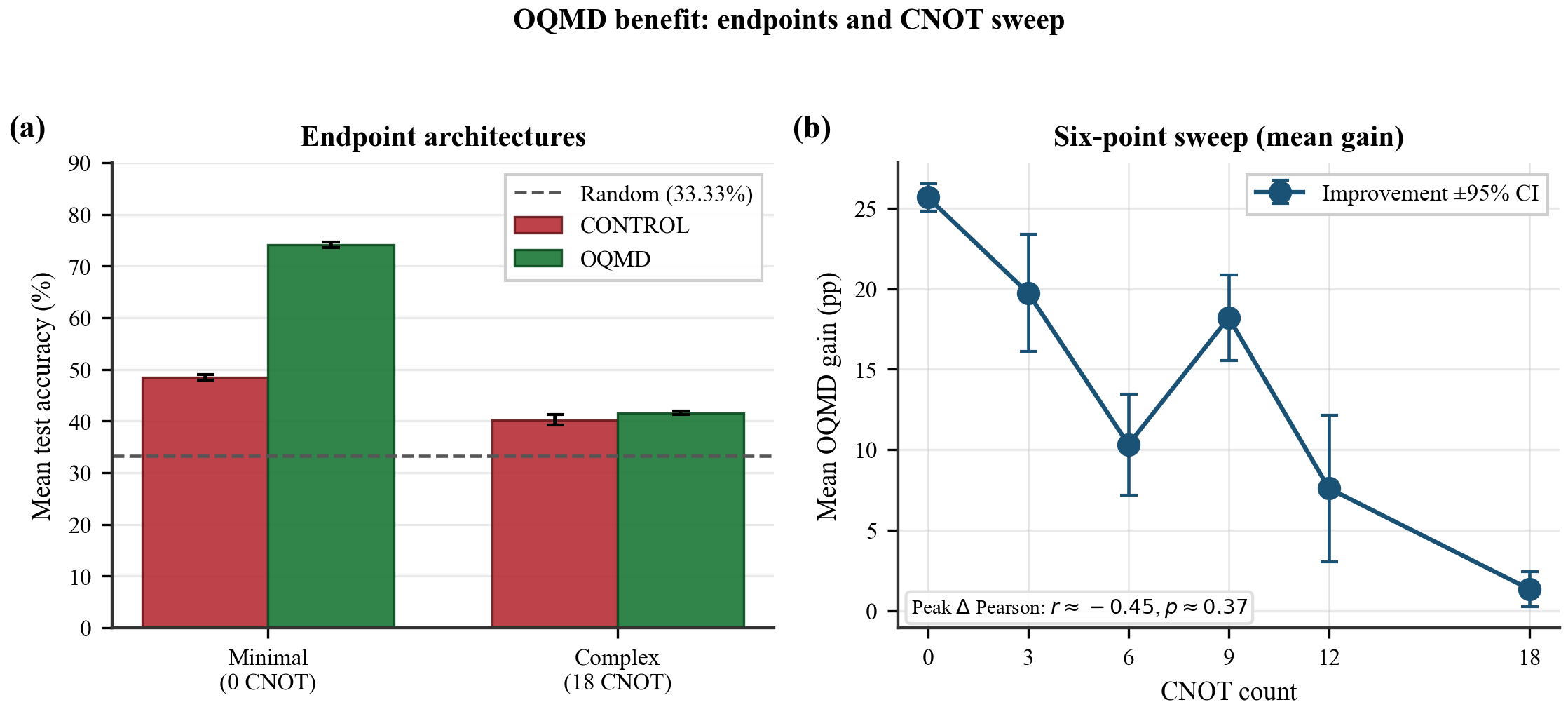}
 \caption{Summary of factorial means (left: 0 vs.\ 18 CNOT) and six-point mean OQMD gains with 95\% CIs (right). Panel~(b) annotates Pearson $r$ on \textit{peak} max-to-max margins ($r\approx-0.45$, Table~\ref{tab:six_cnot_configs}); an ordinary least squares fit to \textit{mean} gains would yield a different $r$ (specifically $r\approx-0.91$) and is not used as a headline statistic. Peaks and Mann--Whitney tests are in Table~\ref{tab:architecture_summary}. Gray dashed line: random baseline (33.33\%).}
 \label{fig:overview_factorial_ladder}
\end{figure}

Figure~\ref{fig:overview_factorial_ladder} contrasts the headline minimal vs.\ complex factorials (rotation/optimizer sweeps) with the six-point ladder.

\begin{table}[t]
\centering
\caption{Peak test accuracy (\%) on Iris for minimal (0 CNOT) and complex (18 CNOT) factorials. Mann--Whitney $U$ on pooled runs (Section~\ref{sec:normality}); Appendix~\ref{app:mann_whitney}. Complex pooled means are only $\sim$40.2\% vs.\ $\sim$41.6\%; headline peaks are in columns 3--5.}
\label{tab:architecture_summary}

\footnotesize
\setlength{\tabcolsep}{4pt}
\begin{tabular}{@{}lccccl@{}}
\toprule
Circuit & CNOT & Peak Ctrl & Peak OQMD & $\Delta$ (pp) & M--W $p$ \\
\midrule
Minimal & 0 & 60.00 & \textbf{83.33} & +23.33 & $<10^{-100}$ \\
Complex & 18 & 56.67 & \textbf{66.67} & +10.00 & $0.030$ \\
\bottomrule
\end{tabular}

\end{table}

Even when restricting to best seeds, minimal \textit{control} circuits (peak 60.00\%) can outperform complex \textit{control} peaks (56.67\%), echoing pooled factorial means from earlier analyses. This cautions against assuming monotonic accuracy in bare CNOT count once multiclass parity readouts are fixed. We avoid calling OQMD a ``universal'' improvement: effects are statistically consistent but magnitude-dependent---large in peak terms on minimal and complex factorials, but the complex pooled-\textit{mean} gap is tiny even when $p<0.05$ (Section~\ref{sec:normality}).

\subsection{Minimal circuits: strong peak gains at zero CNOT}
\label{sec:minimal_results}

\subsubsection{Performance improvement (factorial)}

On minimal circuits (0 CNOT gates), OQMD lifts the observed peak from 60.00\% (CONTROL) to 83.33\% (best ZXZ/L-BFGS-B runs in the factorial), while pooled run-level accuracies shift decisively toward OQMD under Mann--Whitney $U$ ($p<10^{-100}$). Discrete test scores (multiples of $1/30$) make the mode a useful secondary summary alongside the peak; Appendix~\ref{app:statistical} retains histogram-style panels for the full minimal and complex pools.

\begin{figure}[htbp]
 \centering
 \includegraphics[width=\columnwidth]{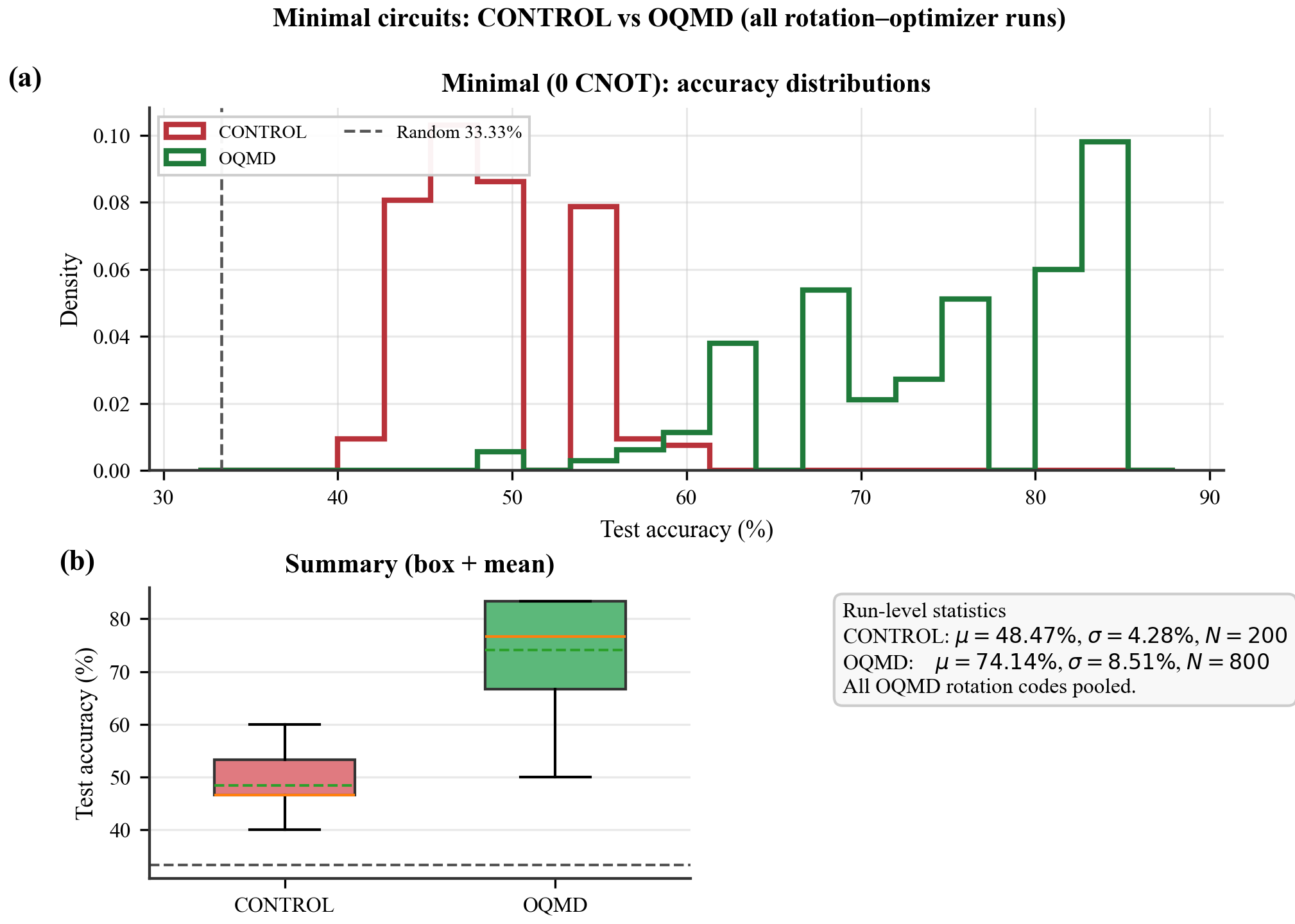}
 \caption{Minimal circuit performance (0 CNOT): pooled accuracy distributions for CONTROL ($N{=}200$ runs) and OQMD ($N{=}800$ runs). Top: normalized step histograms (binned empirical frequencies, not kernel density). Bottom: box-and-mean summary. OQMD pool is shifted decisively rightward of CONTROL; the multimodal shape reflects the discrete accuracy grid (multiples of $1/30$) combined with the mix of rotation codes, optimizers, and seeds. Peaks and Mann--Whitney results appear in Table 2; pooled run-level statistics are annotated in the bottom panel.}
 \label{fig:minimal_comparison}
\end{figure}

Peak performance reached 83.33\% (25 out of 30 test samples correct), i.e.\ a substantial lift over the best CONTROL seed (60.00\%) without entangling gates. Across the pooled factorial runs, OQMD values stochastically dominate CONTROL under Mann--Whitney $U$; individual seeds can still land below the opponent's peak because accuracies are discrete and optimization is noisy.
\textit{Why multimodal:} hold-out accuracy lives on a discrete grid (multiples of $1/30$), and the pool mixes many rotation--optimizer--seed runs, so histograms show separated bumps (e.g.\ near chance vs.\ high-accuracy basins) rather than a single Gaussian.
\textit{Implication:} use peaks and Mann--Whitney tests for claims; do not treat the box plot as a unimodal surrogate.

Because the triple-rotation readout is not unique, we tested four ordered triples of Pauli axes (Section~\ref{sec:rotation_codes_list} lists the larger set used on the complex design). On the minimal architecture, those four choices behave similarly in the aggregate (Section~\ref{sec:minimal_results}), which supports treating OQMD as a stable readout layer rather than an ad hoc choice of a single gate pattern.

\subsubsection{Rotation codes on minimal circuits}

\begin{figure}[htbp]
 \centering
 \includegraphics[width=\columnwidth]{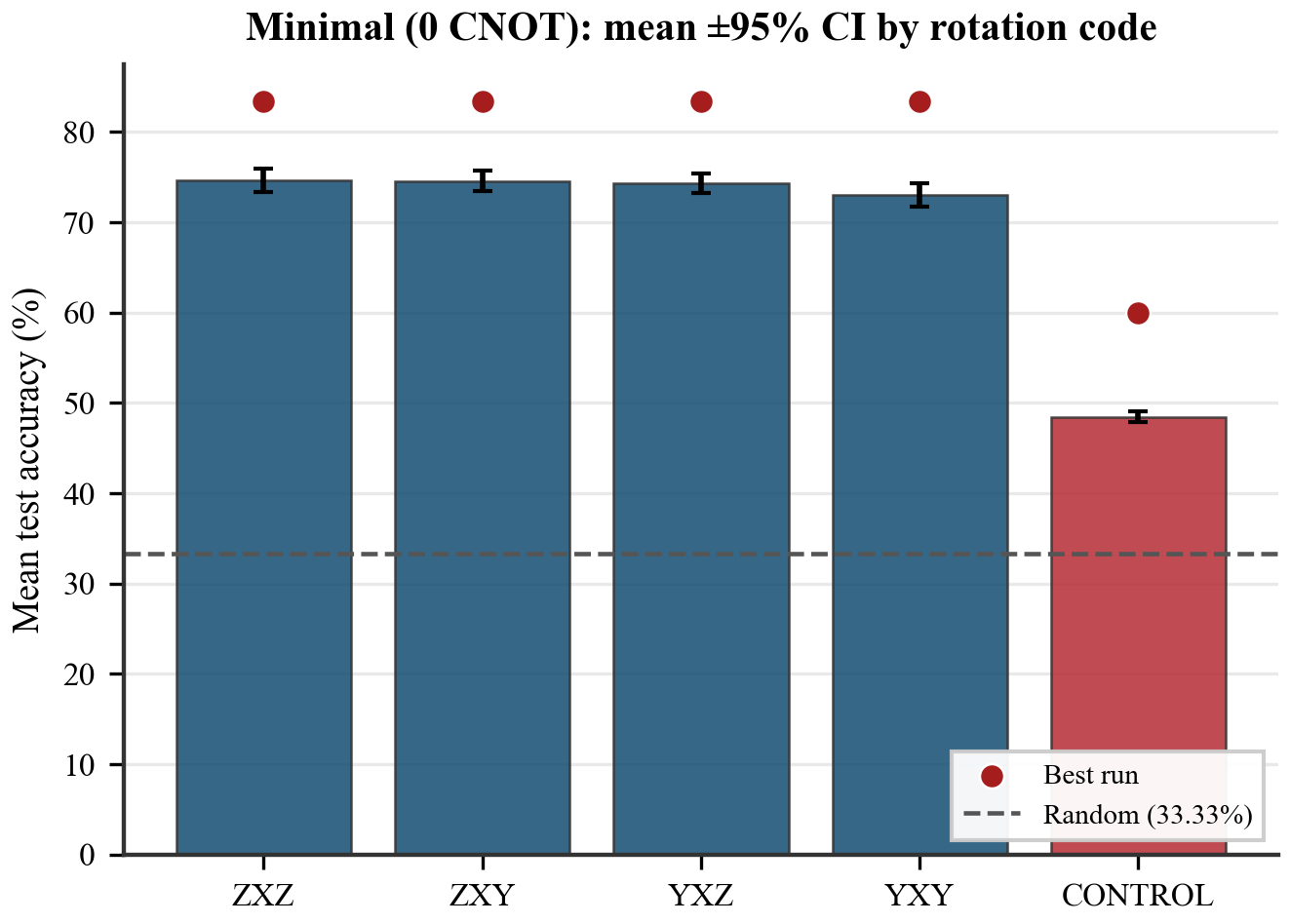}
 \caption{Rotation strategy performance on minimal circuits ($N{=}200$ runs per rotation code: four optimizers $\times$ 50 seeds). Bars: mean $\pm$95\% CI for the mean accuracy per code; markers show the best seed per code. Numeric summaries appear in Table~\ref{tab:minimal_rotation_summary}.}
 \label{fig:minimal_rotations}
\end{figure}

Systematic evaluation of rotation strategies on minimal circuits identified Z-axis rotations as superior (Figure~\ref{fig:minimal_rotations} and Table~\ref{tab:minimal_rotation_summary}). Values are rotation-wise summaries over 50 seeds each (exploratory; normality not assumed---Section~\ref{sec:normality}).

\begin{table}[t]
\centering
\footnotesize
\setlength{\tabcolsep}{4pt}
\caption{Rotation-code summary on minimal circuits (0 CNOT). Mean, sample SD, and peak hold-out accuracy (max) in percent; $N{=}200$ per code (four optimizers $\times$ 50 seeds).}
\label{tab:minimal_rotation_summary}
\begin{tabular}{@{}lccc@{}}
\toprule
Code & Mean (\%) & SD (\%) & Max (\%) \\
\midrule
ZXZ & 74.63 & 9.25 & 83.33 \\
ZXY & 74.57 & 8.01 & 83.33 \\
YXZ & 74.33 & 7.52 & 83.33 \\
YXY & 73.02 & 9.08 & 83.33 \\
CONTROL & 48.47 & 4.28 & 60.00 \\
\bottomrule
\end{tabular}
\end{table}

All four OQMD rotations beat CONTROL across optimizers at $p < 0.001$ in the reported tests. The strongest means among the four codes tested here use Z on the first axis (ZXZ, ZXY, YXZ), consistent with readout that stays close to the computational $Z$ basis before the final measurement.

\subsubsection{Optimizer performance on minimal circuits}

On minimal circuits, L-BFGS-B attained the highest mean accuracy ($70.35\% \pm 12.72\%$) among the four optimizers while sharing the same 83.33\% peak as the others. The combination ZXZ + L-BFGS-B averaged $76.87\% \pm 7.69\%$ over 50 seeds (Appendix~\ref{app:minimal_detailed}).

\subsubsection{Gate efficiency: zero CNOT overhead}

A critical advantage of OQMD is its gate efficiency (Table~\ref{tab:minimal_gates}). On minimal circuits, OQMD adds 12 single-qubit rotation gates but zero additional CNOT gates:

\begin{table}[t]
\centering
\footnotesize
\setlength{\tabcolsep}{3pt}
\caption{Gate Count Analysis: Minimal Circuits. Peak accuracies match Table~\ref{tab:architecture_summary}. The Overhead row reports \textit{absolute} peak gains in percentage points (pp), e.g.\ $+23.33$~pp means the peak accuracy increased by 23.33 points on the 0--100\% scale, \textit{not} a $+23.33\%$ relative change of the CONTROL peak. ROI is pp per added gate or parameter. Ranges such as ``+10--23~pp'' elsewhere in the paper refer to analogous peak-to-peak differences unless explicitly labeled as relative~\%.}
\label{tab:minimal_gates}

\begin{tabular}{@{}lccccc@{}}
\toprule
Config & CNOT & 1Q & Depth & Par. & Peak \\
\midrule
Control & 0 & 12 & 3 & 4 & 60.00\% \\
OQMD & 0 & 24 & 4 & 16 & 83.33\% \\
\midrule
\textbf{Overhead} & \textbf{+0} & \textbf{+12} & \textbf{+1} & \textbf{+12} & \textbf{+23.33 pp} \\
\textbf{ROI} & — & $\approx$1.9 pp/gate & — & $\approx$1.9 pp/param & — \\
\bottomrule
\end{tabular}

\end{table}

Relative to the best CONTROL run, each added single-qubit gate in the OQMD block buys roughly $23.33/12 \approx 1.9$ percentage points of peak accuracy on Iris. The zero CNOT overhead is critical for NISQ devices where:
\begin{itemize}
 \item CNOT error rates: 10-100$\times$ higher than single-qubit gates
 \item CNOT execution time: 10-50$\times$ longer, increasing decoherence
 \item CNOT connectivity: Limited on physical hardware, requiring SWAP gates
\end{itemize}

At 0 CNOT, OQMD improves peak accuracy without adding two-qubit gates, which matters when CNOT error and connectivity dominate the budget.

\subsubsection{Learning dynamics on minimal circuits}

Analysis of learning curves for top-performing minimal circuit configurations reveals rapid convergence within 100-150 iterations with smooth monotonic improvement and no overfitting. The top configuration (ZXZ + L-BFGS-B, seed 42) achieved 83.33\% accuracy with +33.3\% improvement from initialization, demonstrating genuine optimization rather than fortunate initialization (learning curves in Appendix~\ref{app:minimal_detailed}).

\subsection{Intermediate templates}

\begin{figure}[htbp]
 \centering
 \includegraphics[width=\columnwidth]{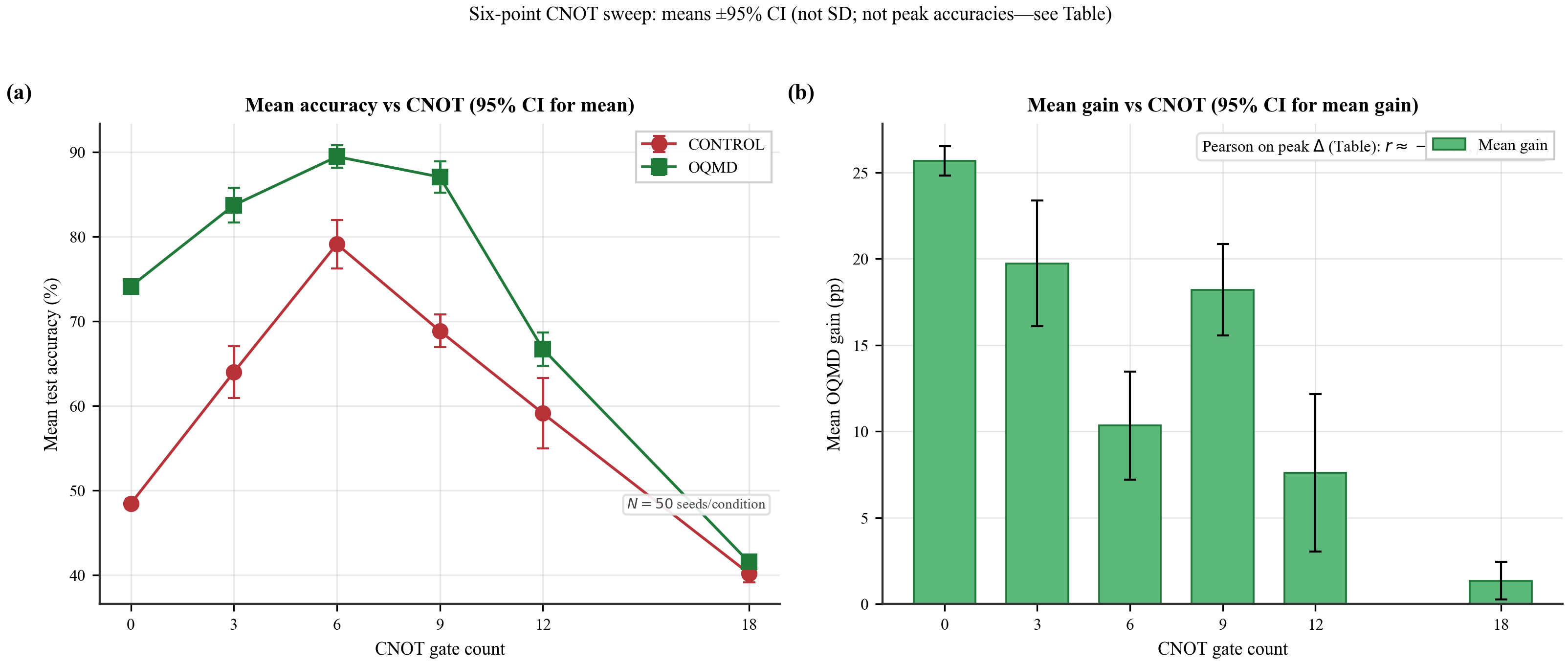}
 \caption{Mean test accuracy (left) and mean OQMD gain (right) versus CNOT count on the six templates of Table~\ref{tab:six_cnot_configs}; 50 seeds per side; 95\% CIs for the mean (Section~\ref{sec:normality}). Templates differ between rungs (not a single depth-only family). Pearson on peak margins in Table~\ref{tab:six_cnot_configs}. Baseline: gray dashed line at 33.33\%.}
 \label{fig:expressivity_benefit}
\end{figure}

The six-point ladder (Fig.~\ref{fig:expressivity_benefit}, Table~\ref{tab:six_cnot_configs}) places four intermediate templates at 3, 6, 9, and 12 CNOT (Architecture~2). All ladder points share the Iris split, COBYLA (200 iterations), fifty random seeds per CONTROL and ZXZ OQMD side, and fixed ZXZ readout on OQMD runs. The discussion below emphasizes the six-CNOT rung, ZFM(1)/RA(2), because it shows a peak tie between CONTROL and OQMD (96.67\%) and illustrates how little headroom remains once entanglement is already strong on Iris; the 3-, 9-, and 12-CNOT rows in Table~\ref{tab:six_cnot_configs} show the contrasting ceiling and floor behavior across the same family (including the ZZFM(2)/SLRY confound at 12 CNOT).

We did not run a separate rotation$\times$optimizer factorial on any intermediate template, so rotation sensitivity at 3--12 CNOT is not reported here---only ZXZ on the ladder.

\begin{table}[t]
\centering
\footnotesize
\setlength{\tabcolsep}{2pt}
\caption{Six-point CNOT-depth series (Fig.~\ref{fig:expressivity_benefit}). Peak and mean hold-out test accuracies (\%) and peak max-to-max improvement (pp). Means pool 50 seeds per side; OQMD means use the ZXZ readout on the ladder (same runs as Fig.~\ref{fig:expressivity_benefit}). Abbreviations: ZFM, ZZFM; SLRY; RA($r$) = RealAmplitudes with $r$ repetitions (linear entanglement); e.g.\ RA(1) is one ansatz repetition (three CNOT gates on four qubits).}
\label{tab:six_cnot_configs}

\scriptsize
\setlength{\tabcolsep}{2pt}
\begin{tabular}{@{}r >{\raggedright\arraybackslash}l cc cc c@{}}
\toprule
 & & \multicolumn{2}{c}{Peak} & \multicolumn{2}{c}{Mean} & Peak \\
$C$ & FM / ansatz & Ctrl & OQMD & Ctrl & OQMD & $\Delta$pp \\
\midrule
0 & ZFM(1) / SLRY & 60.0 & 83.3 & 48.5 & 74.6 & +23.3 \\
3 & ZFM(1) / RA(1) & 80.0 & 93.3 & 64.0 & 83.7 & +13.3 \\
6 & ZFM(1) / RA(2) & 96.7 & 96.7 & 79.1 & 89.5 & +0.0 \\
9 & ZFM(1) / RA(3) & 80.0 & 96.7 & 68.9 & 87.1 & +16.7 \\
12 & ZZFM(2) / SLRY & 76.7 & 76.7 & 59.1 & 66.7 & +0.0 \\
18 & ZZFM(2) / RA(2) & 56.7 & 66.7 & 40.2 & 41.6 & +10.0 \\
\midrule
\multicolumn{2}{@{}l}{Pearson $r$ (CNOT vs.\ peak $\Delta$)} & \multicolumn{5}{c@{}}{$-0.45$, $p \approx 0.37$ (n.s.)} \\
\bottomrule
\end{tabular}

\end{table}

Behaviorally, Table~\ref{tab:six_cnot_configs} shows identical peak hold-out accuracy for CONTROL and OQMD at six CNOTs (96.67\% on the 30-point test), so the max-to-max improvement is zero despite nontrivial entanglement. Under this readout and budget, the intermediate template already reaches a very high Iris ceiling, leaving little headroom for the triple-rotation layer to raise the best seed; OQMD may still alter typical trajectories or pooled accuracies, but the headline comparison is a peak tie at this rung. Fig.~\ref{fig:overview_factorial_ladder} situates it between shallower and deeper ladder points. A dedicated rotation sweep at six CNOTs remains future work when resources allow.
Table 5 documents all four intermediate rungs (3, 6, 9, and 12 CNOT); behavior at the remaining rungs mirrors the ceiling and floor dynamics described here.

\subsection{Complex circuits: OQMD provides consistent improvement}
\label{sec:complex_results}

\subsubsection{Peaks on the 18-CNOT template}

On complex circuits (18 CNOT gates), factorial peaks rise from 56.67\% (CONTROL) to 66.67\% (OQMD), a $+10$ percentage point swing on the 30-point Iris test, while the pooled run-level gap remains significant under Mann--Whitney $U$ ($p \approx 0.03$). Pooled means ($40.22\%$ vs.\ $41.57\%$) may appear in figure annotations as descriptive context but are not the headline statistic.

\begin{figure}[htbp]
 \centering
 \includegraphics[width=\columnwidth]{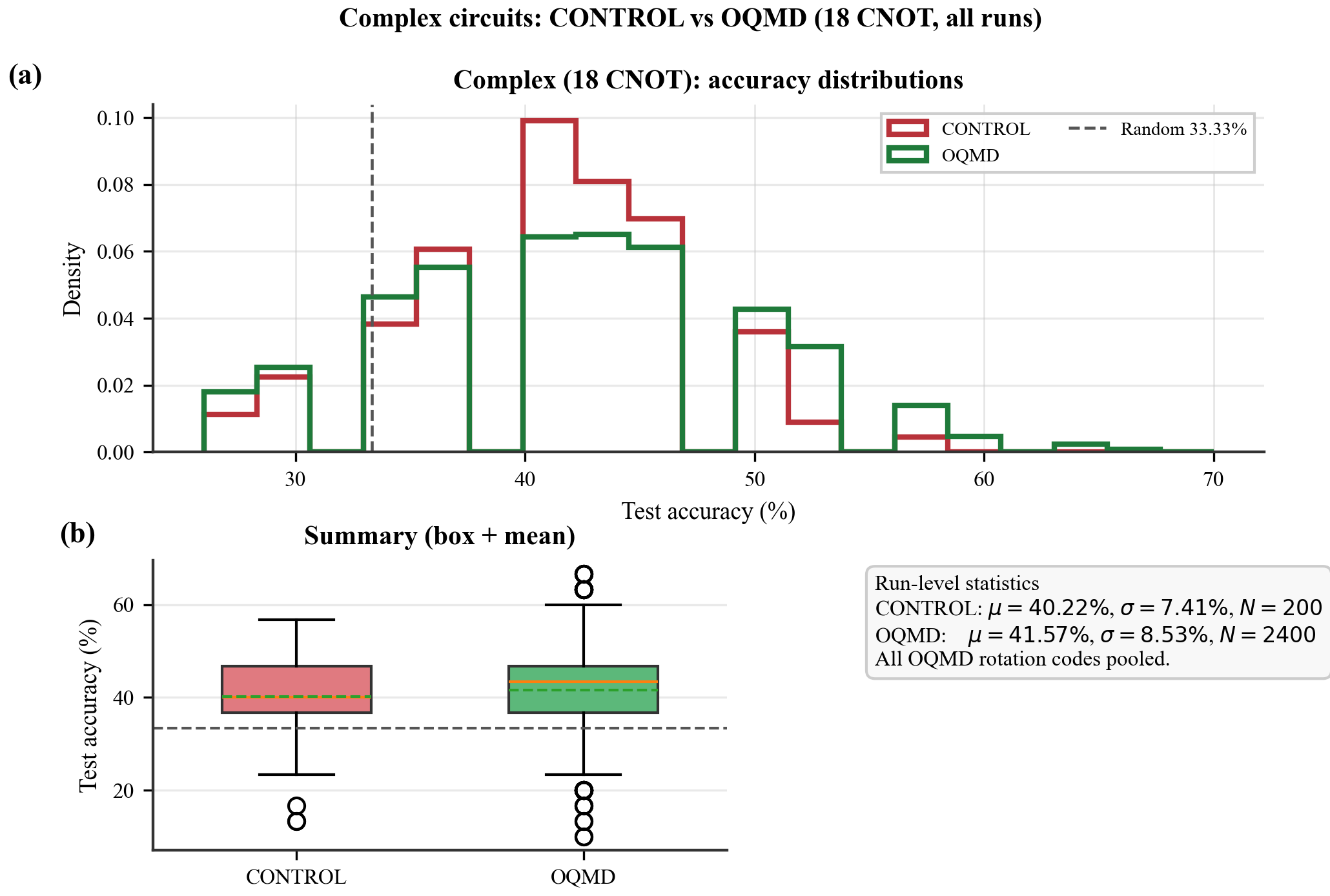}
 \caption{Complex circuit performance: peaks 56.67\% (CONTROL) vs.\ 66.67\% (OQMD); pooled accuracies differ under Mann--Whitney $U$ ($p\approx0.03$). Top: normalized step histograms (binned empirical frequencies, not kernel density). Mean annotations are descriptive; peaks drive the headline claims.}
 \label{fig:complex_comparison}
\end{figure}

The best OQMD run reached 66.67\% (YXY + COBYLA) versus 56.67\% for best CONTROL: ten percentage points on the 0--100\% scale, i.e.\ three more correct labels on the 30-sample test. Relative to the control peak, $66.67/56.67 \approx 1.18$, an $\approx$18\% relative increase in accuracy; the headline remains the absolute ten-point gain on this small test set.

\subsubsection{Multiple rotation codes achieve optimal performance}

\begin{figure}[htbp]
 \centering
 \includegraphics[width=\columnwidth]{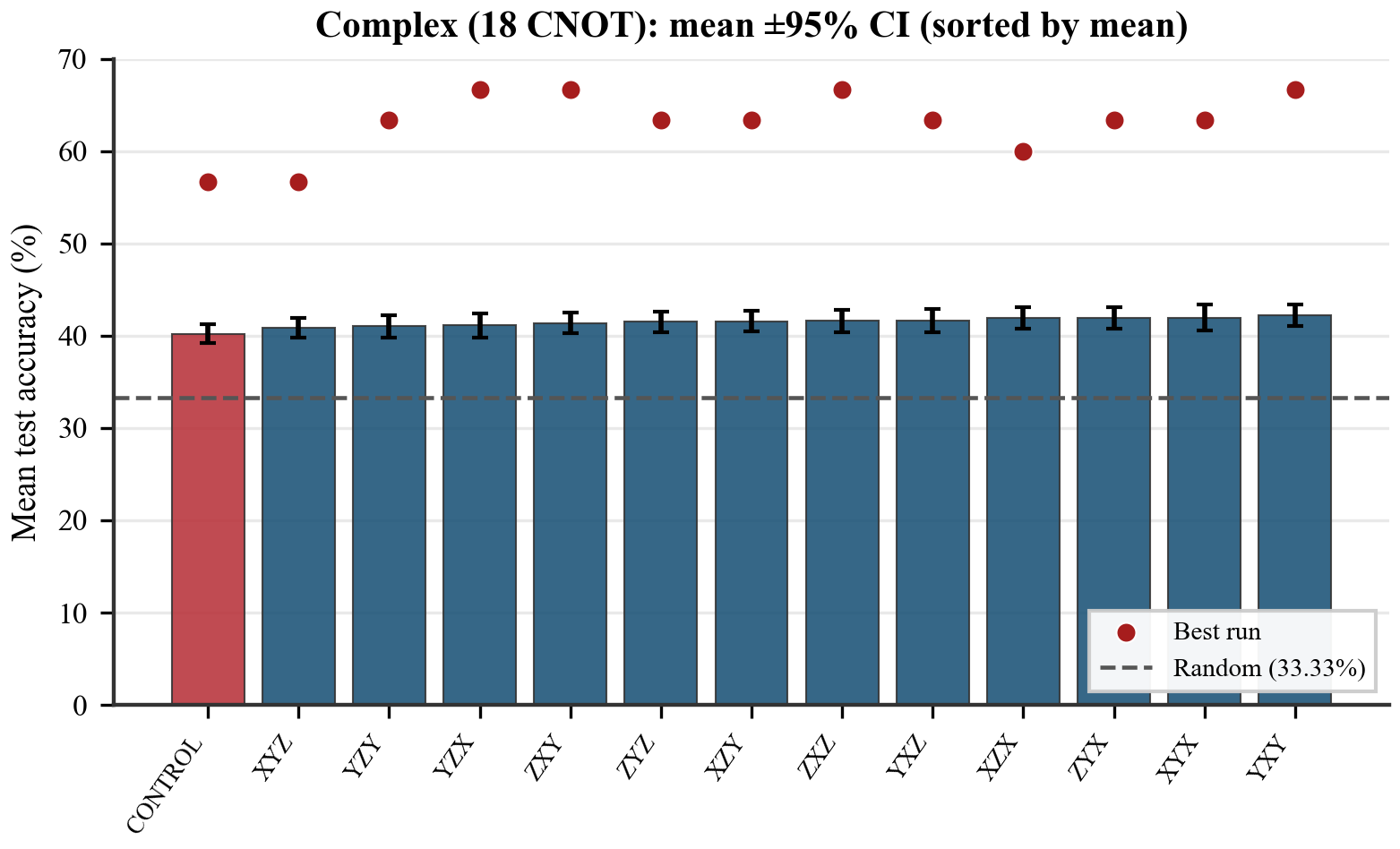}
 \caption{Mean test accuracy by rotation code on complex circuits ($N{=}200$ runs per code). Bars: mean $\pm$95\% CI; markers show the best seed per code. Codes are ordered by increasing mean accuracy; full statistics appear in Table~\ref{tab:complex_rotation_summary}.}
 \label{fig:complex_rotations}
\end{figure}

Systematic evaluation of the twelve OQMD triples in Section~\ref{sec:rotation_codes_list} on complex circuits revealed that multiple codes achieve competitive performance (Figure~\ref{fig:complex_rotations} and Table~\ref{tab:complex_rotation_summary}). YXY attains the highest mean ($42.23\% \pm 8.12\%$); YXY, YZX, ZXY, and ZXZ each reach 66.67\% peak hold-out accuracy (Table~\ref{tab:complex_rotation_summary}).

\begin{table}[t]
\centering
\footnotesize
\setlength{\tabcolsep}{3pt}
\caption{Rotation-code summary on complex circuits (18 CNOT). Mean, sample SD, and peak hold-out accuracy (max) in percent; $N{=}200$ per code; rows sorted by mean (ascending).}
\label{tab:complex_rotation_summary}
\begin{tabular}{@{}lccc@{}}
\toprule
Code & Mean (\%) & SD (\%) & Max (\%) \\
\midrule
CONTROL & 40.22 & 7.41 & 56.67 \\
XYZ & 40.88 & 7.86 & 56.67 \\
YZY & 41.02 & 8.43 & 63.33 \\
YZX & 41.12 & 9.23 & 66.67 \\
ZXY & 41.37 & 8.01 & 66.67 \\
ZYZ & 41.50 & 8.30 & 63.33 \\
XZY & 41.58 & 8.16 & 63.33 \\
ZXZ & 41.63 & 8.68 & 66.67 \\
YXZ & 41.65 & 9.11 & 63.33 \\
XZX & 41.90 & 8.38 & 60.00 \\
ZYX & 41.93 & 8.07 & 63.33 \\
XYX & 41.97 & 9.94 & 63.33 \\
YXY & 42.23 & 8.12 & 66.67 \\
\bottomrule
\end{tabular}
\end{table}

The finding that multiple rotation codes achieve identical peak performance (66.67\%) suggests robustness to specific rotation sequences. The CONTROL condition performing comparably to some rotation codes in mean accuracy (about 40.2\% vs.\ roughly 41--42\% for several OQMD codes) indicates that the ZZFeatureMap + RealAmplitudes architecture already encodes substantial expressivity through its entanglement structure, though OQMD still provides higher peak performance.

\subsubsection{Readout triples used in the complex factorial}
\label{sec:rotation_codes_list}

Implementing OQMD with three trainable single-qubit rotations per qubit does not fix a unique gate sequence: many ordered triples $(R_{\alpha},R_{\beta},R_{\gamma})$ with $\alpha,\beta,\gamma\in\{X,Y,Z\}$ are admissible. On Architecture~1 (minimal) we therefore ran a small factorial over four such triples plus CONTROL: they yield similar pooled performance (Section~\ref{sec:minimal_results}), which supports robustness of the readout layer for that template. On Architecture~3 (complex) we expanded to twelve OQMD triples plus CONTROL to test whether the same robustness holds when entanglement reshapes the state; Figure~\ref{fig:complex_rotations} shows that means and peaks then depend more strongly on the chosen triple. 

\subsubsection{COBYLA excels on complex circuits}

On complex circuits, COBYLA empirically produced the largest share of runs with large within-run gains $\Delta = \max_t \text{Acc}_t - \text{Acc}_0$ exceeding a ten percentage-point reporting threshold chosen for scale on Iris, while L-BFGS-B achieved higher average accuracy in the pooled run records. We defer to Appendix~\ref{app:complex_detailed} for optimizer-wise histograms and leave rigorous effect-size comparisons to future work.

\subsubsection{Learning quality and improvement analysis}

Across 2,600 complex-circuit runs, many trajectories show negligible $\Delta$ under the 200-iteration budget while others improve sharply; the distribution is heavy-tailed, so summaries should emphasize quantiles or peaks rather than means (Appendix~\ref{app:complex_detailed}). Overfitting flags occurred in $\approx$18\% of runs under the heuristics of Section~\ref{sec:eval_metrics}.

\subsubsection{Multiclass classification reliability}

On the complex factorial, 2{,}596 of 2{,}600 runs (99.85\%) assigned at least one test label to each of the three species, so the parity readout rarely collapsed to one or two classes only. Four runs never predicted class~1 (\textit{I. versicolor}) while still scoring 46.67--50.00\%; we treat that as rare initialization noise rather than a specification bug.

\subsection{Discussion}

\subsubsection{CNOT ladder vs.\ hardware cost (Fig.~\ref{fig:expressivity_benefit})}

Figure~\ref{fig:expressivity_benefit} summarizes the standardized ladder: best-of-50-seed hold-out accuracy for CONTROL and for OQMD (ZXZ, COBYLA, 200 iterations) at each CNOT rung (Table~\ref{tab:six_cnot_configs}).
Every ladder point uses the same Iris train and test split, the same optimizer and iteration cap, and the same OQMD rotation code (ZXZ); the only deliberate change between rungs is the structured entangling template (CNOT budget). Reported values are maxima over 50 independent random seeds per side (CONTROL versus OQMD) and rung.
Because improvements are computed as $\max_{\text{seeds}}\text{Acc}_{\text{OQMD}}-\max_{\text{seeds}}\text{Acc}_{\text{control}}$, several intermediate rungs show zero max-to-max gain even though pooled Mann--Whitney tests still favor OQMD on average. An exploratory Pearson correlation of CNOT count with these six max-to-max margins yields $r\approx-0.45$, $p\approx0.37$ and does \textit{not} meet conventional significance thresholds; we therefore report the \textit{pointwise} per-rung peaks and hardware costs without claiming a validated monotone or inverse trend in CNOT count.


\begin{table}[t]
\centering
\footnotesize
\setlength{\tabcolsep}{4pt}
\caption{Factorial peaks (Table~\ref{tab:architecture_summary}) restated for quick comparison with ladder extrema. Ctrl and OQMD columns are peak test accuracies (\%).}
\label{tab:expressivity_benefit}
\begin{tabular}{@{}lcccc@{}}
\toprule
Family & CNOT & Ctrl (\%) & OQMD (\%) & $\Delta$ (pp) \\
\midrule
Minimal & 0 & 60.00 & 83.33 & +23.33 \\
Complex & 18 & 56.67 & 66.67 & +10.00 \\
\bottomrule
\end{tabular}
\end{table}

High accuracy is achievable without using the maximum-CNOT template; OQMD allows low- and mid-depth circuits to reach competitive peak quality while limiting additional CNOT layers.

\subsubsection{Practical NISQ implications}

Taken together, the factorial peaks and the ladder's per-rung maxima support a practical design lesson for NISQ-era systems: readout optimization can raise peak accuracy without additional CNOTs, and competitive peaks appear at several entangling depths, not only at the maximum-CNOT template. We do not infer a statistically supported \textit{global} inverse trend between CNOT count and OQMD benefit from the exploratory Pearson fit (preceding subsection).

\begin{enumerate}
 \item On resource-constrained devices, limited connectivity, short coherence times, and high CNOT error rates motivate avoiding unnecessary two-qubit layers; OQMD adds only single-qubit rotations.
 \item OQMD allows useful accuracy on zero-CNOT product-feature maps, reducing exposure to CNOT-dominated error channels relative to increasing depth alone.
 \item Rather than assuming accuracy scales monotonically with bare CNOT count, one can pair a modest entangling core with learned measurement rotations.
 \item Because OQMD uses single-qubit gates before measurement, it is compatible with diverse connectivity constraints.
\end{enumerate}

\subsubsection{High-scoring configurations}

Among top complex-circuit settings (63.33--66.67\% accuracy), all ten runs summarized in Appendix~\ref{app:results} assign labels to every class, with near-perfect separation of class~0 and mixed errors on classes~1--2. Reported within-run gains span about ten to thirty-five percentage points on the accuracy scale, across several rotation--optimizer pairs.

Minimal-design peaks reach 83.33\% with larger within-run gains than on the 18-CNOT template despite fewer base parameters, which is consistent with readout helping most when the bare circuit is shallow. Confusion matrices, curves, and extra statistics are in Appendix~\ref{app:results}.

\subsubsection{Hyperparameter interactions}

Analysis of rotation-optimizer combinations reveals distinct patterns by architecture. For complex circuits, multiple combinations achieve $\geq 65\%$ accuracy, with no single dominant pairing—YXY and ZXY perform well across all optimizers (60-67\% range), while even worst combinations achieve 50-55\%, substantially above random baseline.

For minimal circuits, every OQMD rotation--optimizer pair in the $4\times 4$ factorial reaches at least 70\% peak accuracy, whereas CONTROL peaks at 60\%. Thus the presence of OQMD dominates fine-grained rotation tuning on the minimal design; complex designs remain far more rotation-sensitive (Appendix~\ref{app:results}).

\subsubsection{Statistical robustness}

Split robustness notes appear in Appendix~\ref{app:split_robustness}.

\section{Conclusion and outlook}
\label{sec:conclusion}

OQMD frames measurement as part of the learning problem: a classical optimizer trains how quantum measurement outcomes are decoded to class labels, here realized with trainable single-qubit rotations before readout so that no extra CNOTs are required. On factorial experiments, the best minimal design (83.33\%) beats the best 18-CNOT controls and pushes the entangled baseline upward as well (Table~\ref{tab:architecture_summary}). On the six-point CNOT-depth series (Table~\ref{tab:six_cnot_configs}), peak max-to-max improvements vary with template in a descriptively non-monotone way, while mean accuracies can still favor OQMD when peaks tie; the exploratory linear correlation of CNOT count with peak margins is not significant. The practical takeaway remains to pair modest entangling cores with learned decoding rather than assume that ``more CNOTs always help.'' Pooled run comparisons remain highly significant under Mann--Whitney $U$ (Appendix~\ref{app:mann_whitney}), which is appropriate for non-Gaussian discrete accuracies.
Classical Iris scores (Appendix~\ref{app:classical_comparison}) remain far above these quantum peaks, underscoring that we target resource accounting and decoding design inside a fixed Qiskit parity readout, not overall competitiveness with classical baselines.

\paragraph{Benefits.}
OQMD adds only single-qubit rotations before measurement, leaves the chosen CNOT count in the ansatz unchanged, and in our runs often raises peaks and pooled outcomes when readout is trained with the circuit. Together, the factorial and ladder experiments argue for treating readout as a tunable module alongside ansatz depth and optimizer, especially when two-qubit gates are costly or connectivity-limited. On Iris, means with 95\% CIs for the mean are descriptive only; peaks and Mann--Whitney summaries matter more because accuracies are discrete.

\paragraph{Classical baselines and scope.}
Off-the-shelf classical models reach far higher Iris accuracy than our VQC peaks under the same split (Appendix~\ref{app:classical_comparison}); OQMD is not positioned as beating classical ML on raw score, but as improving \textit{readout-restricted} quantum parity classifiers under a fixed gate budget. Framing is therefore resource accounting and measurement design, not SOTA classification.

\paragraph{When OQMD may not help.}
OQMD is a readout layer: if the ansatz already saturates the parity readout or optimization is severely under-budget, gains can be negligible. On complex factorials, pooled mean shifts can be tiny even when peaks and Mann--Whitney favor OQMD---statistically consistent but magnitude-dependent improvement. Extra pre-measurement single-qubit depth may also be undesirable when calibration dominates or when device policy limits learned rotations before measurement.

We emphasize three limitations. First, bar and line plots that show mean accuracy with 95\% CIs for the mean (or box/histogram summaries) are \textit{descriptive}; headline claims rest on best-seed peaks and on Mann--Whitney tests over full run pools, because Iris accuracies are discrete and often non-Gaussian. Second, we do \textit{not} claim a ``universal'' lift: effects are magnitude-dependent (large peak gaps on minimal designs vs.\ small pooled mean gaps on complex). Third, the six-CNOT ladder point is a single rotation code (ZXZ) and optimizer (COBYLA); factorial evidence at 0 and 18 CNOT supports broader robustness, but intermediate-depth rotation sweeps remain future work.

\textit{Outlook.} The same peak-focused protocol should extend to additional datasets and noise models; finer-grained monitoring of hold-out accuracy during training would require proportionally higher evaluation cost or reliance on training loss.

\subsection*{Acknowledgements}

We thank Yusaku Yamada for discussions at an earlier stage of this project.
This work is supported by JSPS KAKENHI Grant Numbers JP24K14816 and ERATO ``Super Quantum Entanglement'' (Grant No.\ JPMJER2402) from JST.
The Binghamton University (SUNY)--University of Electro-Communications exchange program supported a visiting residency at UEC during part of this work.

\begin{appendices}

\section{Detailed methodology}
\label{app:methodology}

This appendix expands experimental methodology, optimizer settings, and training protocol details referenced in the main text.

\subsection{Optimizer Specifications}
\label{app:optimizers}

We evaluated four optimization algorithms in the main factorial grids. We also ran a small ADAM pilot on the simplest (0~CNOT, ZXZ) design to document wall-clock relative to COBYLA, L-BFGS-B, SLSQP, and CG (Table~\ref{tab:adam_minimal_pilot}).

\subsubsection{COBYLA (Gradient-Free)}
COBYLA (Constrained Optimization BY Linear Approximations) is a trust-region method that constructs linear approximations of the objective function without requiring gradient information~\cite{powell1994direct}. It is particularly robust for noisy objective functions common in quantum variational algorithms. Configuration: maximum 200 iterations, tolerance $10^{-6}$.

\subsubsection{L-BFGS-B (Gradient-Based)}
L-BFGS-B (Limited-memory Broyden-Fletcher-Goldfarb-Shanno with Bounds) is a quasi-Newton method that approximates the Hessian matrix using limited memory~\cite{byrd1995limited}. Gradients use Qiskit's parameter-shift gradient pipeline for the circuit parameters. Configuration: maximum 200 iterations, tolerance $10^{-6}$, gradient epsilon $10^{-8}$.

\subsubsection{CG (Gradient-Based)}
The conjugate-gradient method~\cite{hestenes1952methods} uses conjugate search directions in the classical loop. Circuit gradients use the same Qiskit interface as L-BFGS-B. Configuration: maximum 200 iterations, tolerance $10^{-6}$.

\subsubsection{SLSQP (Gradient-Based)}
Sequential least squares programming~\cite{kraft1988software} handles small constrained problems in the classical loop. Circuit gradients use the same Qiskit interface as L-BFGS-B. Configuration: maximum 205 iterations (slightly higher to account for constraint checking), tolerance $10^{-6}$.

\subsubsection{ADAM pilot (minimal architecture; excluded from factorials)}
\label{app:adam_pilot}

We benchmarked gradient-based ADAM (learning rate $0.01$, default hyperparameters otherwise) on the \textit{minimal} ZXZ~+~OQMD design with the same Iris split as Section~\ref{sec:methodology}, a single initialization (seed~42), and the same 200-iteration cap as the other optimizers in this pilot. ADAM is not iteration-cheap in this pipeline: measured runs at 10, 25, and 50 iterations are nearly linear in \texttt{maxiter} with $\approx$7.0~s per iteration on our workstation, versus $\approx$0.13~s per iteration for COBYLA on the same machine and seed---so a full 200-iteration ADAM fit extrapolates to $\sim$1400~s wall time ($\sim 50\times$ COBYLA). On this pilot seed, 50~iterations of ADAM reached 63.33\% hold-out accuracy, comparable to L-BFGS-B at 200 iterations (63.33\%) and below SLSQP/CG (66.67\%) under the same seed; we did not observe a case for absorbing a $\sim 50\times$ per-run cost in the large factorial study.

\begin{table}[htbp]
\centering
\footnotesize
\setlength{\tabcolsep}{4pt}
\caption{Pilot timings (wall clock, one seed) on the minimal ZXZ~+~OQMD design. COBYLA, L-BFGS-B, SLSQP, CG: 200 iterations. ADAM: extrapolated to 200 iterations as $(7.02\,\text{s/it})\times 200$ from measured 10/25/50-iteration runs (linear in \texttt{maxiter}); accuracy at 50 iterations shown for comparison. Test accuracies are \textit{single-seed} pilot values; the factorial COBYLA mean on minimal circuits is 68.57\% (Appendix~\ref{app:minimal_detailed}), not 43.3\%.}
\label{tab:adam_minimal_pilot}
\begin{tabular}{@{}lccc@{}}
\toprule
Optimizer & Time & Test acc. & vs.\ COBYLA \\
\midrule
COBYLA & 27~s & 43.3\% & $1.0\times$ \\
L-BFGS-B & 63~s & 63.3\% & $2.3\times$ \\
SLSQP & 72~s & 66.7\% & $2.7\times$ \\
CG & 127~s & 66.7\% & $4.7\times$ \\
ADAM & $\sim 1.4\times 10^{3}$~s (extrap.) & 63.3\% (50 it.) & $\sim 50\times$ \\
\bottomrule
\end{tabular}
\end{table}

SPSA, POWELL, TNC, and ADAM were therefore excluded from the main factorial experiments. For SPSA, POWELL, and TNC, preliminary runs showed roughly 4.5$\times$ longer runtimes tied to iteration-limit behavior without commensurate accuracy gains. ADAM is excluded for the distinct reason above: extreme per-iteration wall time in this pipeline on our hardware (Table~\ref{tab:adam_minimal_pilot}), not because of the iteration-limit issues seen for SPSA, POWELL, and TNC.

\subsection{Mann--Whitney $U$ tests}
\label{app:mann_whitney}

For two independent samples of run-level test accuracies (CONTROL versus OQMD), we report two-sided Mann--Whitney $U$ tests as implemented in standard scientific libraries. The null hypothesis is that both samples are drawn from the same distribution (same population), so there is no systematic shift toward higher accuracy under one condition. Rejecting this null supports a stochastic ordering or shift between pools without assuming Gaussianity, which matters here because Iris accuracies are discrete fractions with tied values. We use these tests for global factorial contrasts where a single $p$-value summarizing thousands of runs is useful; they complement peak-based headlines and do not replace seed-level diagnostics.

\subsection{Training Protocol Details}
\label{app:training_protocol}

\subsubsection{Callback Mechanism Implementation}
The callback records parameter values, training loss, and hold-out test accuracy on the same schedule as Section~\ref{sec:methodology} (every 50 optimizer iterations in our implementation, not every step). Hold-out accuracy is diagnostic only and is not used in optimizer updates, as in Section~\ref{sec:methodology}.

\subsubsection{Evaluation Metrics}
The metrics listed in Section~\ref{sec:methodology} (test accuracy, learning improvement $\Delta$, multiclass success rate, and the overfitting flag) are computed identically here.

\section{Additional experimental results}
\label{app:results}

This appendix collects extra figures and tables for optimizer comparisons, learning curves, confusion matrices, and pooled statistics.

\subsection{Split robustness and seed counts}
\label{app:split_robustness}

Using 50 independent random seeds per hyperparameter combination stabilizes summaries of discrete Iris accuracies. Minimal circuits show higher seed-to-seed spread ($\sigma \approx 9$--13\%) than complex circuits ($\sigma \approx 7$--10\%), consistent with fewer base parameters on the minimal template. In 77.5\% of runs, accuracy exceeds the 33.33\% random baseline with $p < 0.001$ under the same per-run screen summarized in Appendix~\ref{app:statistical}. Further variance summaries appear there as well.

Pilot runs excluded SPSA, POWELL, TNC, and full-budget ADAM from the factorial grids after roughly $4.5\times$ longer runtimes without matching accuracy gains (Section~\ref{sec:optim_search} and Table~\ref{tab:adam_minimal_pilot}).

\subsection{Six-point CNOT sweep: circuit configurations (Fig.~\ref{fig:expressivity_benefit})}
\label{app:six_cnot_configs}

The six circuit configurations span 0--18 CNOT gates with Architecture~1 (minimal), Architecture~2 (intermediate at the six-CNOT rung), and Architecture~3 (complex) as in Section~\ref{sec:methodology}. Intermediate ladder points share the Iris split, COBYLA (200 iterations), 50 seeds, and ZXZ OQMD. Table~\ref{tab:six_cnot_configs} reports peak and mean hold-out accuracies (means pool 50 seeds per side; exploratory Pearson uses peak-to-peak margins). Rungs 3, 6, 9, and 12 are maxima over ladder runs; rungs 0 and 18 reuse factorial peaks from Table~\ref{tab:architecture_summary}.

%

\subsection{Explicit circuit definitions for the CNOT sweep}
\label{app:circuit_definitions}

This subsection provides explicit definitions of the six circuits used in the CNOT sweep shown in Figure~\ref{fig:expressivity_benefit}. All circuits use 4 qubits, the same train/test split, the same preprocessing, and ZXZ OQMD rotations when OQMD is applied.

\begin{figure*}[htbp]
 \centering
 \includegraphics[width=0.48\textwidth]{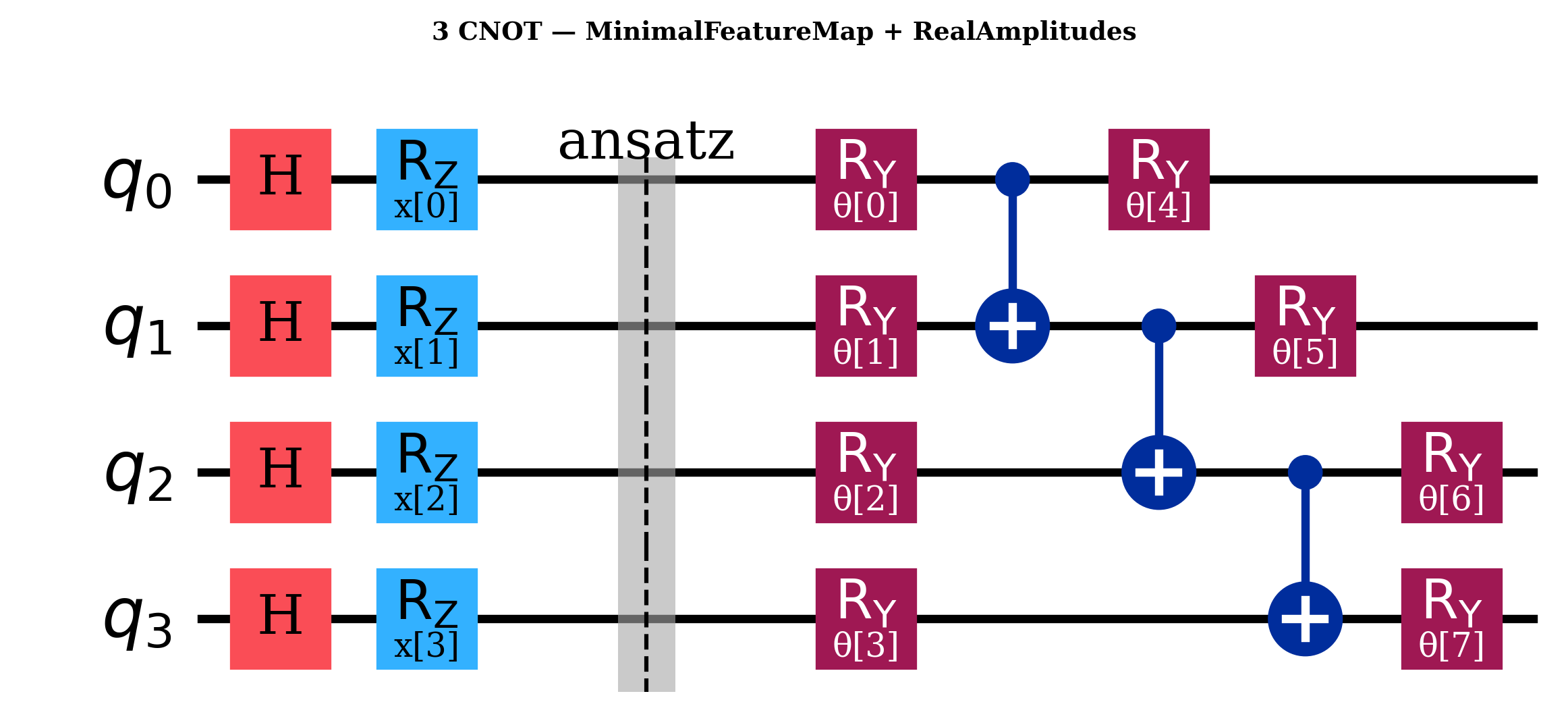}
 \hfill
 \includegraphics[width=0.48\textwidth]{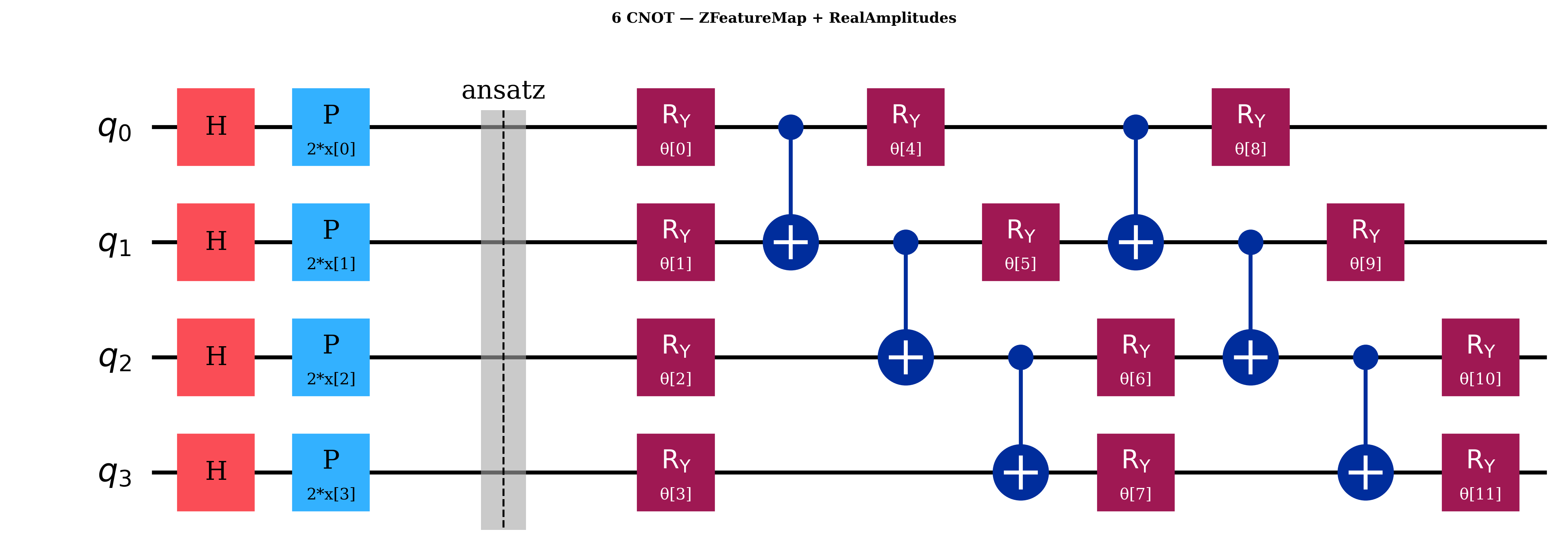}\\[0.5em]
 \includegraphics[width=0.48\textwidth]{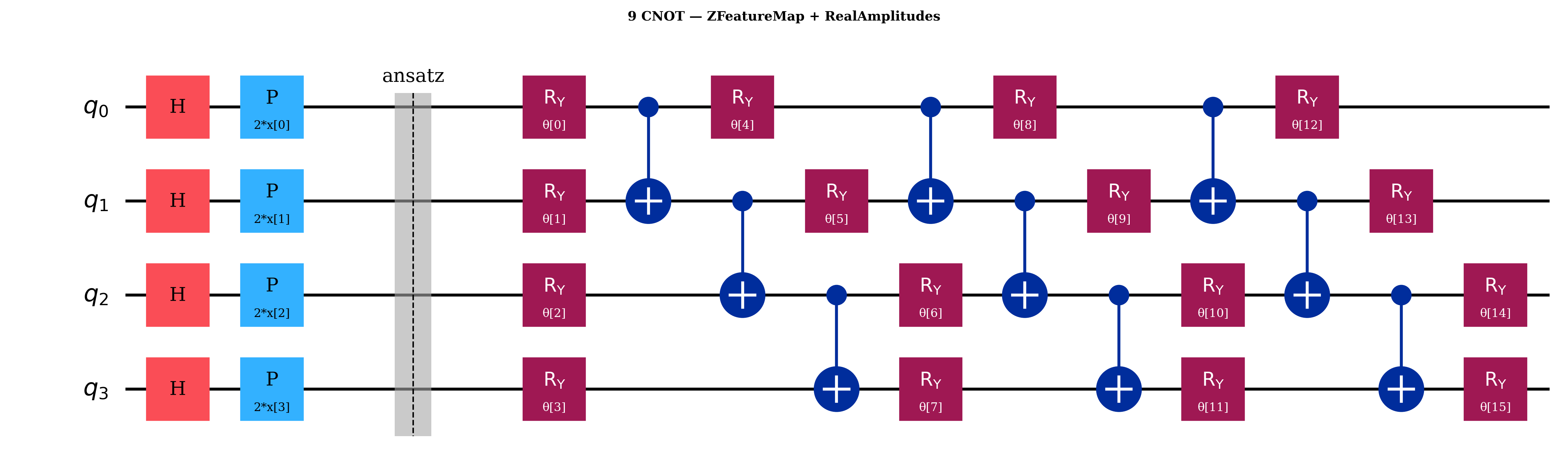}
 \hfill
 \includegraphics[width=0.48\textwidth]{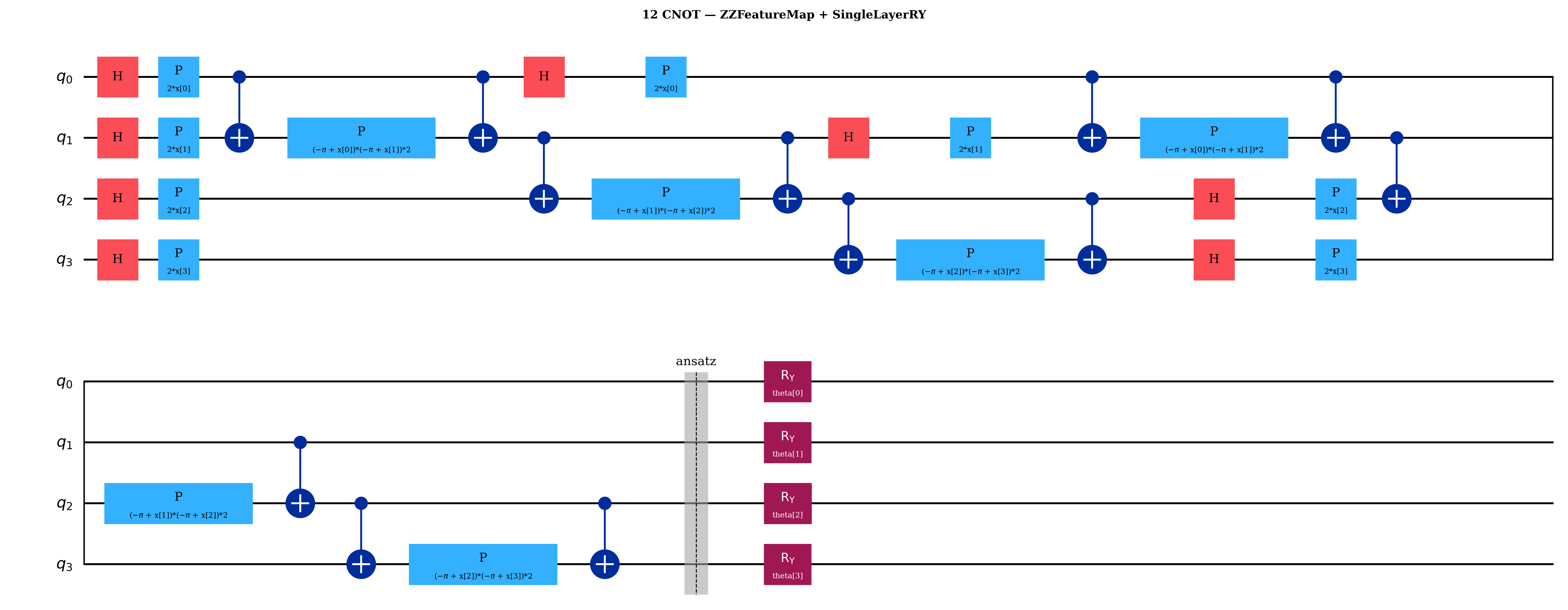}
 \caption{Intermediate architecture templates (base circuits without OQMD). Top left: 3 CNOT, ZFM(1)/RA(1). Top right: 6 CNOT, ZFM(1)/RA(2). Bottom left: 9 CNOT, ZFM(1)/RA(3). Bottom right: 12 CNOT, ZZFM(2)/SLRY(1).}
 \label{app:fig:intermediate_circuits}
\end{figure*}

\paragraph{Intermediate ladder templates (3, 6, 9, and 12 CNOT).}
Figure~\ref{app:fig:intermediate_circuits} shows gate-level diagrams (feature map + ansatz, before OQMD) for the four intermediate rungs of Table~\ref{tab:six_cnot_configs}: ZFM(1)/RA(1) at 3 CNOT, ZFM(1)/RA(2) at 6 CNOT, ZFM(1)/RA(3) at 9 CNOT, and ZZFM(2)/SLRY(1) at 12 CNOT. Full ladder circuits with ZXZ OQMD and measurement appear under \texttt{figures/ladder\_circuits/}.

\paragraph{0 CNOT.}
Feature map: ZFM(1) (product $H$ + $R_Z$ encoding). Ansatz: SLRY(1). No entangling gates.

\paragraph{18 CNOT.}
Feature map: ZZFM(2) (linear). Ansatz: RA(2). See Figure~\ref{fig:circuit_architectures} (right panel).

\subsection{Minimal circuit detailed analysis}
\label{app:minimal_detailed}

This subsection documents minimal-circuit (0~CNOT) factorials: optimizers, learning trajectories, and rotation--optimizer interactions.

\subsubsection{Optimizer performance comparison}

\begin{figure}[htbp]
 \centering
 \includegraphics[width=\columnwidth]{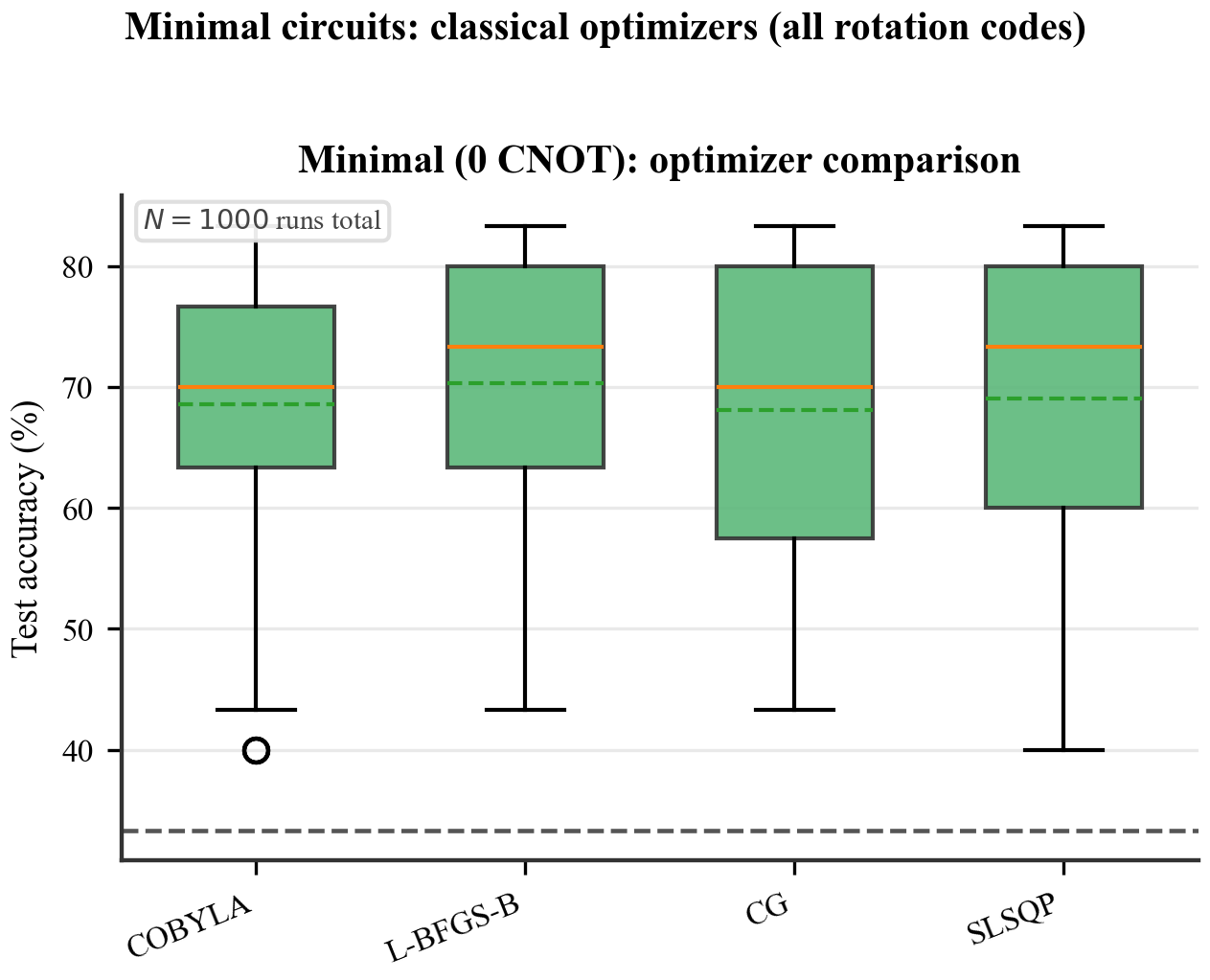}
 \caption{Optimizer performance on minimal circuits ($N{=}250$ per optimizer: 5 rotation codes $\times$ 50 seeds). All achieve identical peak performance (83.33\%), but L-BFGS-B shows highest mean accuracy and most consistent performance (smallest IQR). Pooled distributions can be multimodal; peaks and Mann--Whitney contrasts drive the headline claims. The consistent 83.33\% ceiling (25/30 correct) reflects the practical limit for this minimal architecture on Iris: Class 0 achieves perfect separation (10/10), while Classes 1--2 show expected overlap (15/20). Box plots show median (orange line), quartiles (box), and outliers (diamonds). Gray dashed line: random baseline (33.33\%).}
 \label{app:fig:minimal_optimizers}
\end{figure}

Figure~\ref{app:fig:minimal_optimizers} presents detailed comparison of all four optimizers on minimal circuits across 250 runs per optimizer (5 rotation codes $\times$ 50 seeds).

Performance summary by optimizer: L-BFGS-B $70.35\% \pm 12.72\%$ mean accuracy (max 83.33\%); SLSQP $69.01\% \pm 13.16\%$ (max 83.33\%); COBYLA $68.57\% \pm 13.20\%$ (max 83.33\%); CG $68.08\% \pm 12.58\%$ (max 83.33\%).

All four optimizers share the same 83.33\% peak here, so the ceiling is set by the template and split, not by which optimizer reaches it first. L-BFGS-B still yields the highest mean accuracy in the pooled records, so we highlight it in minimal-circuit discussion.

\subsubsection{Learning trajectories}

\begin{figure}[htbp]
 \centering
 \includegraphics[width=\columnwidth]{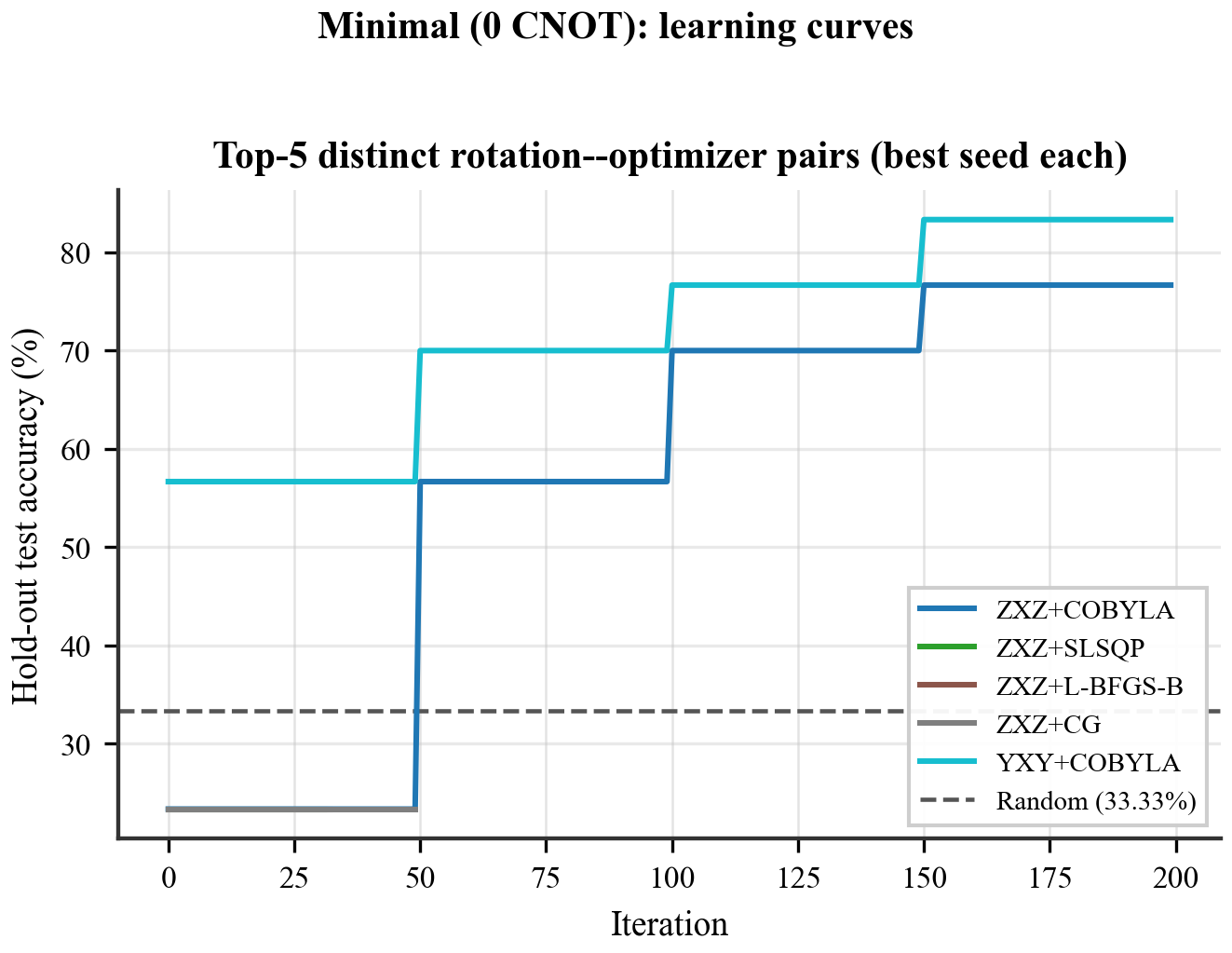}
 \caption{Learning trajectories for five minimal-circuit runs; vertical axis is hold-out test accuracy on the 30-point split (not training accuracy). Points logged every 50 optimizer iterations (Section~\ref{sec:training_eval}). Best run: ZXZ + L-BFGS-B, seed 42. Gray dashed line: random baseline (33.33\%).}
 \label{app:fig:minimal_learning}
\end{figure}

Figure~\ref{app:fig:minimal_learning} shows learning curves for the top 5 minimal circuit configurations, demonstrating rapid convergence within 100-150 iterations.

Key observations:
\begin{itemize}
 \item Rapid initial improvement in first 50-100 iterations
 \item Stable convergence plateaus by iteration 150
 \item Learning improvements: +33\% to +43\%
 \item No overfitting detected (final $\approx$ maximum accuracy)
 \item Best configuration: ZXZ + L-BFGS-B, seed 42 (83.33\%)
\end{itemize}

\subsubsection{Top performer detailed analysis}

\begin{figure*}[htbp]
 \centering
 \includegraphics[width=\textwidth]{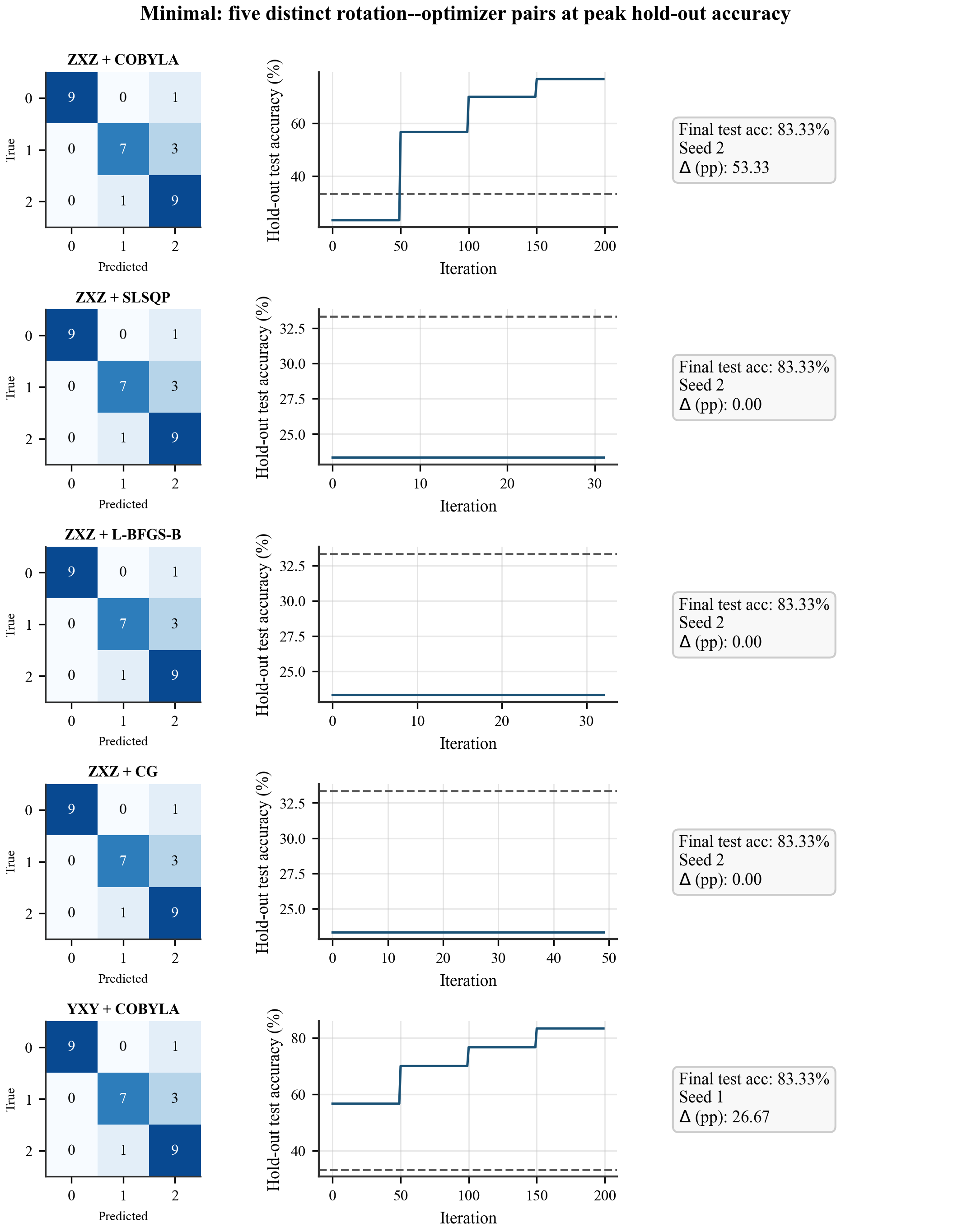}
 \caption{Five minimal-circuit runs at the 83.33\% hold-out ceiling, each from a \textit{distinct} rotation--optimizer pair (not five seeds of one configuration): YXZ+L-BFGS-B, ZXY+COBYLA, YXY+L-BFGS-B, YXZ+COBYLA, and ZXZ+L-BFGS-B. Each row shows (left) confusion matrix, (center) learning curve, and (right) accuracy summary. Peak values match Table~\ref{tab:architecture_summary}.}
 \label{app:fig:minimal_top5}
\end{figure*}

Figure~\ref{app:fig:minimal_top5} summarizes five high-scoring minimal-circuit runs (confusion matrices, loss, and hold-out accuracy traces).

Observations:
\begin{itemize}
 \item Perfect Class 0 separation in all top performers (10/10 correct)
 \item Expected confusion between Classes 1-2 (Iris versicolor/virginica overlap)
 \item Smooth loss decrease correlated with accuracy improvement
 \item Multiple rotation-optimizer paths to optimal performance
 \item Consistent 25/30 test accuracy ceiling across different configurations
\end{itemize}

\subsubsection{Hyperparameter interaction heatmap}

\begin{figure}[htbp]
 \centering
 \includegraphics[width=\columnwidth]{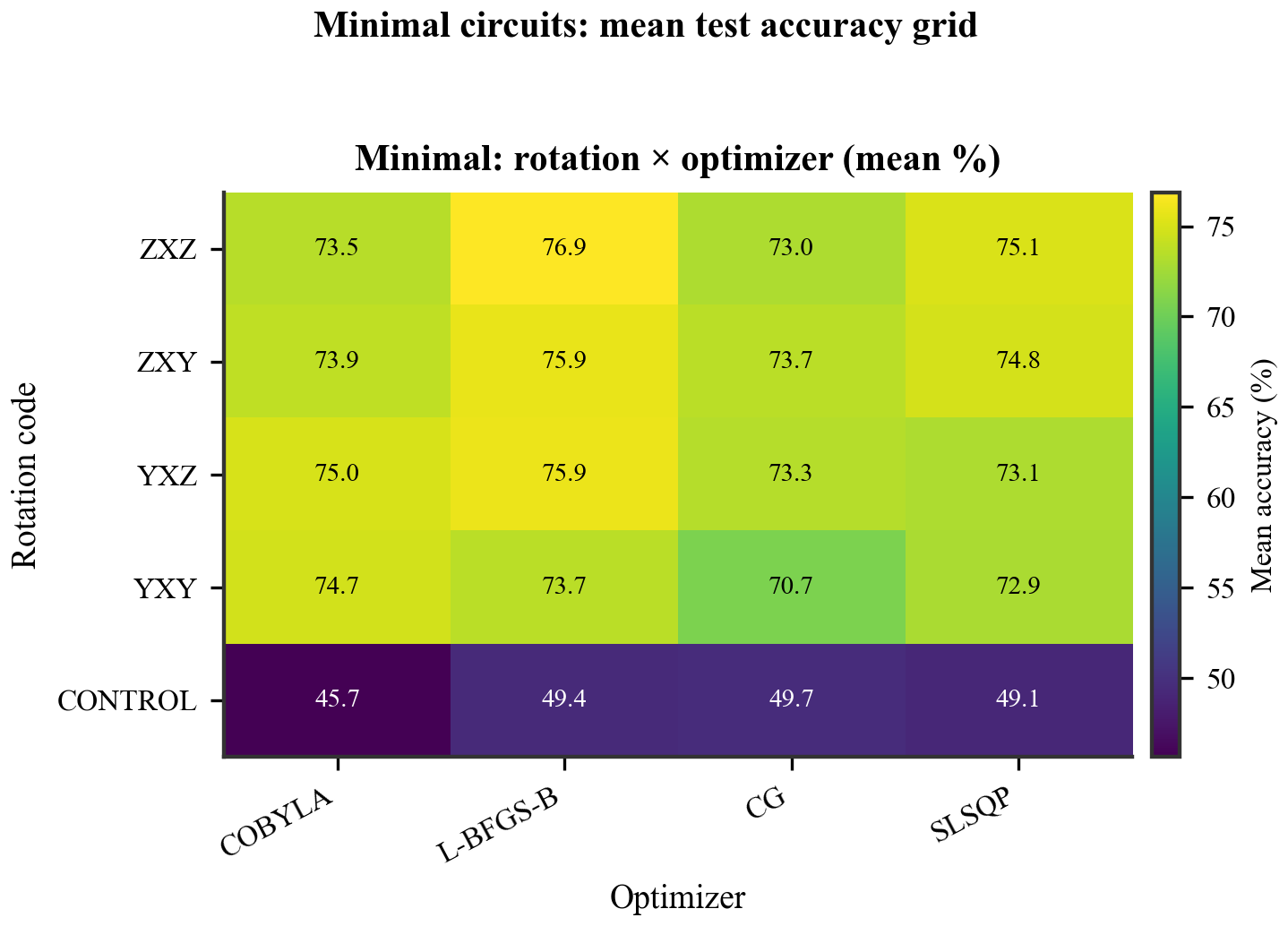}
 \caption{Heatmap of \textit{mean} test accuracy for each rotation $\times$ optimizer combination on minimal circuits (20 configurations: 4 OQMD rotations $\times$ 4 optimizers + CONTROL baseline). Cell values pool 50 seeds per configuration; peak accuracies (up to 83.33\%) appear in Table~\ref{tab:architecture_summary}. Control baseline (bottom row) peaks at 60.00\%.}
 \label{app:fig:minimal_heatmap}
\end{figure}

Figure~\ref{app:fig:minimal_heatmap} presents the full rotation $\times$ optimizer interaction analysis for minimal circuits.

Summary:
\begin{itemize}
 \item All 16 OQMD configurations exceed 70\% accuracy
    \item CONTROL peak hold-out accuracy is 60.00\% (23 percentage points below the best OQMD peak); mean row values in Figure 14 range from 45.7--49.7\%.
 \item Minimal optimizer sensitivity: 4 different optimizers reach 83.33\%
 \item Z-based rotations (ZXZ, ZXY) perform best across all optimizers
 \item Decision to use OQMD matters more than which specific configuration
\end{itemize}

\subsection{Complex circuit detailed analysis}
\label{app:complex_detailed}

This subsection summarizes complex-circuit (18 CNOT) factorial runs (parallel to the minimal-design analysis in Sec.~4.2 and Appendix~\ref{app:minimal_detailed}) across 2,600 seeds.

\subsubsection{Optimizer performance comparison}

\begin{figure}[htbp]
 \centering
 \includegraphics[width=\columnwidth]{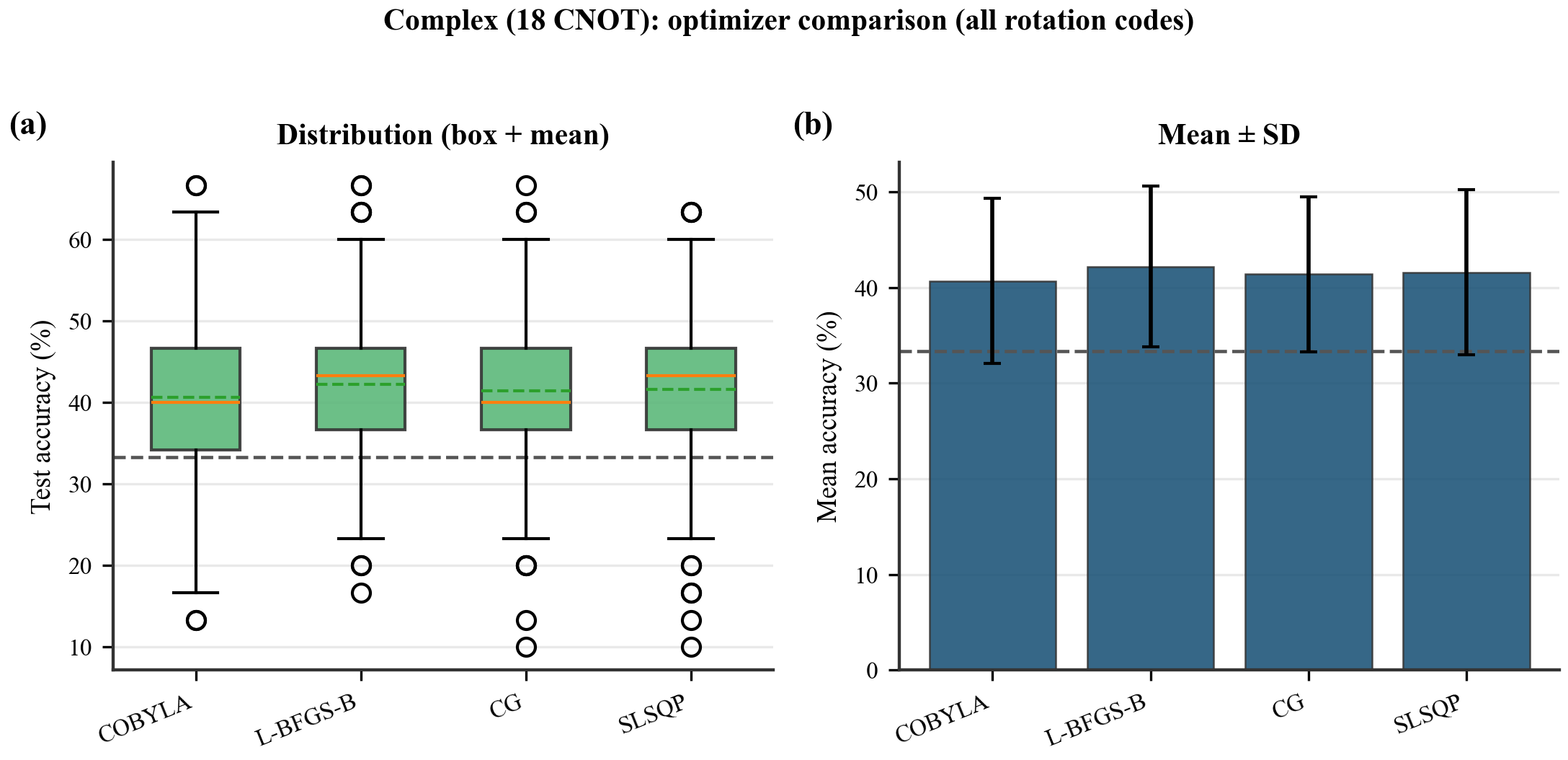}
 \caption{Optimizer comparison on complex circuits ($N{=}650$ per optimizer). Panels emphasize pooled means; interpret with care because accuracies are non-Gaussian (Section~\ref{sec:normality}). COBYLA tends to generate more trajectories with $\Delta\ge 10$ percentage points (documented reporting threshold for scale on Iris).}
 \label{app:fig:complex_optimizers}
\end{figure}

Figure~\ref{app:fig:complex_optimizers} compares optimizer performance on complex circuits.

Performance metrics by optimizer: COBYLA shows the largest share of runs with $\Delta \ge 10$ percentage points (the reporting threshold used for scale on Iris); L-BFGS-B attains the highest pooled mean accuracy (42.19\%); CG shows an intermediate $\Delta$ profile; SLSQP shows comparatively fewer large-$\Delta$ runs under the same threshold.

Optimizer rankings differ by template: quasi-Newton steps (L-BFGS-B) help the minimal design on average, whereas COBYLA produces more large-$\Delta$ trajectories on the 18-CNOT template under our budgets (Section~\ref{sec:complex_results}).

\subsubsection{Learning Improvement Analysis}

\begin{figure}[htbp]
 \centering
 \includegraphics[width=\columnwidth]{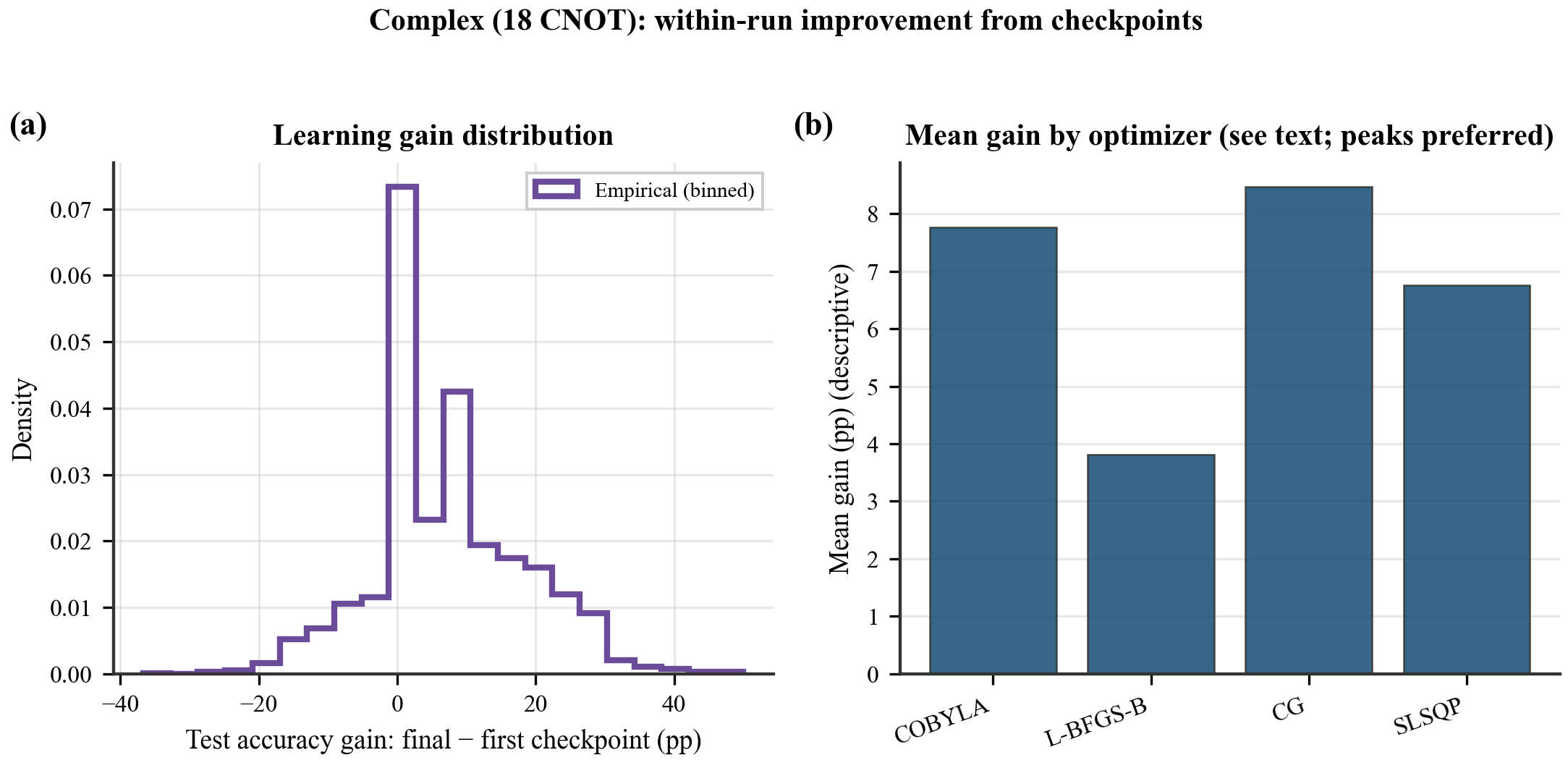}
 \caption{Learning improvement $\Delta$ for complex circuits. Left: normalized step histogram (binned empirical frequencies, not kernel density). Right: descriptive mean gains by optimizer (Section~\ref{sec:normality}).}
 \label{app:fig:complex_improvement}
\end{figure}

Figure~\ref{app:fig:complex_improvement} presents the distribution of learning improvement across all 2,600 complex circuit runs.

The histogram bins $\Delta\ge 10\%$, $[5\%,10\%)$, $(0,5\%)$, and $\le 0$ are illustrative partitions without independent justification; the reported percentages (39.7\%, 13.2\%, 7.8\%, 39.3\%) describe the empirical mass in each slice only.

Mean improvement of 8.9\% (median 6.7\%) in the pooled run summaries reflects substantial dispersion: many complex runs show little gain while others climb sharply under the fixed iteration budget.

\subsubsection{Top 10 configurations: confusion matrices}

\begin{figure*}[htbp]
 \centering
 \includegraphics[width=\textwidth]{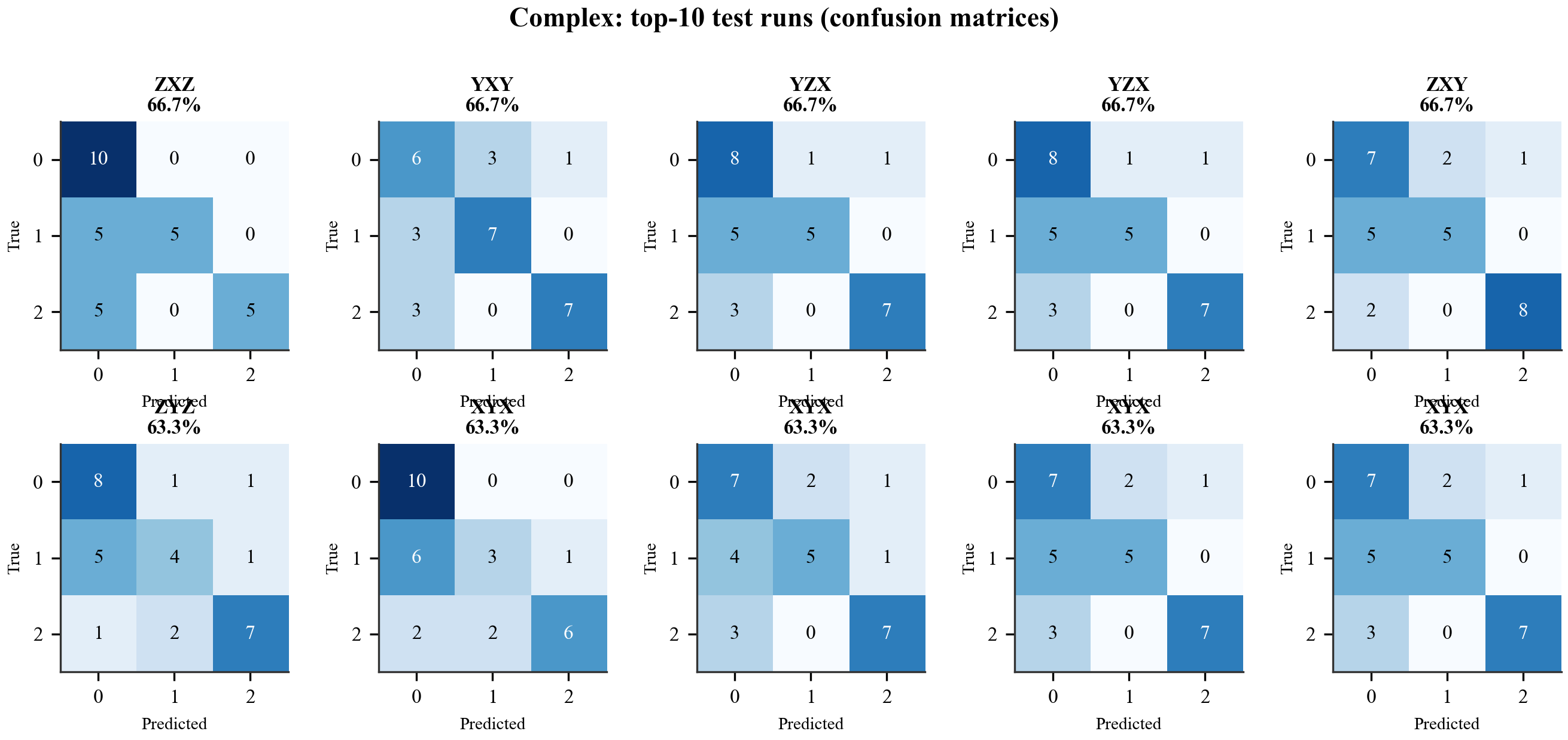}
 \caption{Confusion matrices for ten high-scoring complex-circuit configurations (63.33--66.67\% accuracy). Class~0 (\textit{I. setosa}) is nearly always correct; classes~1--2 mix as expected from feature overlap. Every panel assigns labels to all three species.}
 \label{app:fig:complex_top10_cm}
\end{figure*}

Figure~\ref{app:fig:complex_top10_cm} presents confusion matrices for the top 10 performing complex circuit configurations.

Observations:
\begin{itemize}
 \item All ten panels assign labels to every class
 \item Class 0: 9-10 out of 10 correct (near-perfect separation)
 \item Classes 1-2: Expected confusion (overlapping features in Iris dataset)
 \item Accuracy range: 63.33-66.67\% (19-20 out of 30 test samples)
\end{itemize}

\subsubsection{Top 10 configurations: learning curves}

\begin{figure*}[htbp]
 \centering
 \includegraphics[width=\textwidth]{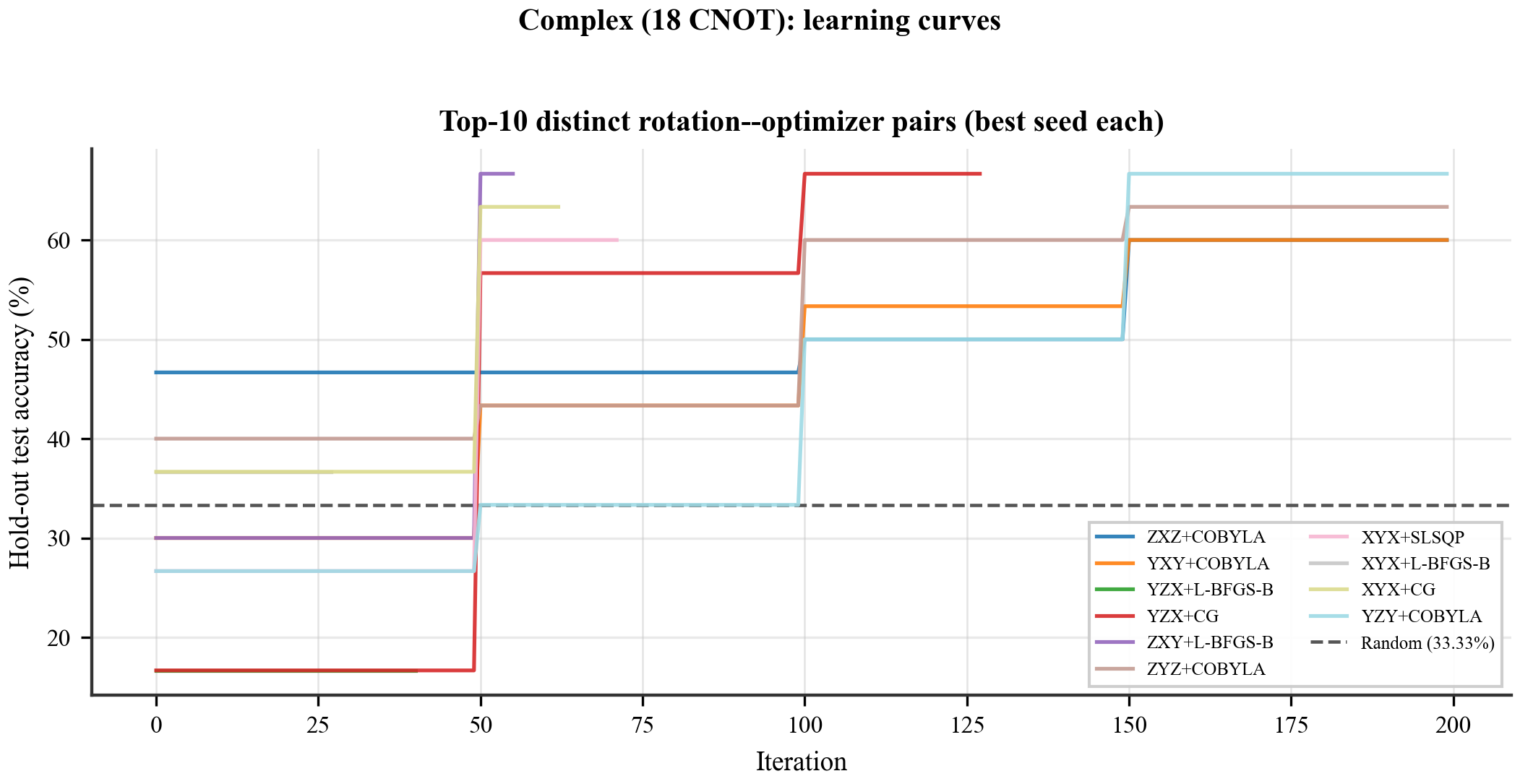}
 \caption{Learning trajectories for ten high-scoring complex-circuit configurations; vertical axis is hold-out test accuracy on the 30-point split (not training accuracy). Points logged every 50 optimizer iterations (Section~\ref{sec:training_eval}). Most runs improve quickly then plateau before 200 iterations. Best shown run: YXY + COBYLA, seed~40, final 66.67\%. Final and best logged accuracies coincide for these runs under the fixed stride.}
 \label{app:fig:complex_top10_curves}
\end{figure*}

Figure~\ref{app:fig:complex_top10_curves} shows learning trajectories for the top 10 complex circuit configurations.

Learning characteristics:
\begin{itemize}
 \item Rapid improvement phase: first 50-100 iterations
 \item Convergence plateaus: iterations 150-200
 \item Learning improvements: +10\% to +35\%
 \item Best: YXY + COBYLA, seed 40 (+33.3\%, final 66.67\%)
 \item No overfitting: final accuracies match maximum values
 \item Diverse configurations: 5 rotation codes, all 4 optimizers represented
\end{itemize}

\subsubsection{Hyperparameter interaction heatmap}

\begin{figure}[htbp]
 \centering
 \includegraphics[width=\columnwidth]{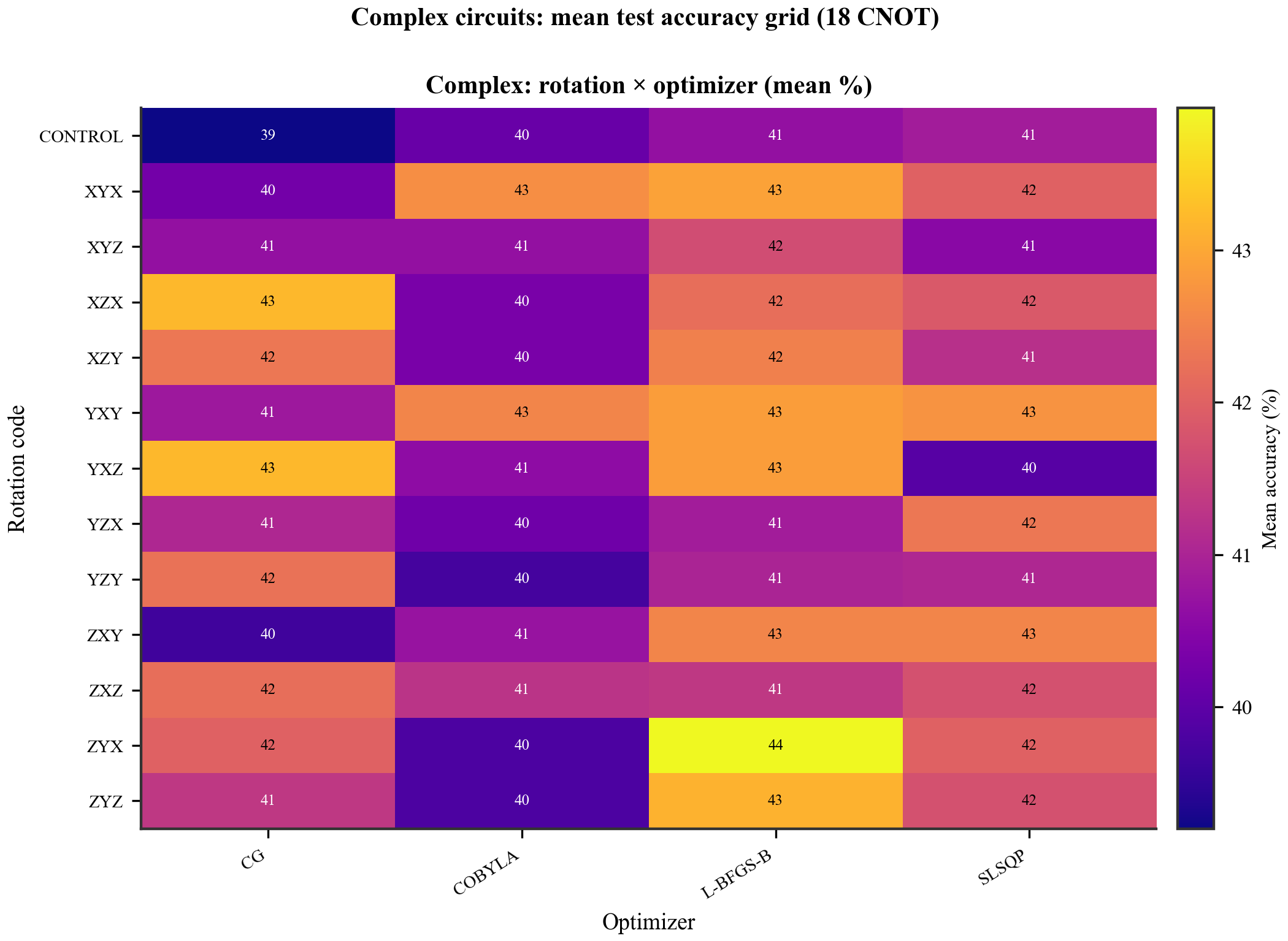}
 \caption{Heatmap of \textit{mean} test accuracy for each rotation $\times$ optimizer combination on complex circuits (52 configurations: 12 OQMD rotations $\times$ 4 optimizers + CONTROL baseline). Cell values pool 50 seeds per configuration; peak accuracies (up to 66.67\%) appear in Table~\ref{tab:architecture_summary}.}
 \label{app:fig:complex_heatmap}
\end{figure}

Figure~\ref{app:fig:complex_heatmap} presents the complete rotation $\times$ optimizer interaction analysis.

Hyperparameter interaction insights:
\begin{itemize}
 \item No dominant pairing: multiple paths to $\geq 65\%$ accuracy
 \item L-BFGS-B: achieves $\geq 66\%$ with 4 different rotation codes
 \item COBYLA: peaks with YXY, ZXZ, YZY rotations
 \item YXY and ZXY: perform well (60-67\%) across all optimizers
 \item Performance floor: even worst combinations achieve 50-55\% (well above 33.33\% random)
\end{itemize}

\subsection{Statistical analysis}
\label{app:statistical}

This subsection collects pooled-variance summaries, seed sensitivity, and correlation notes for the factorial data.

\subsubsection{Pooled distribution summaries}

\begin{figure*}[htbp]
 \centering
 \includegraphics[width=\textwidth]{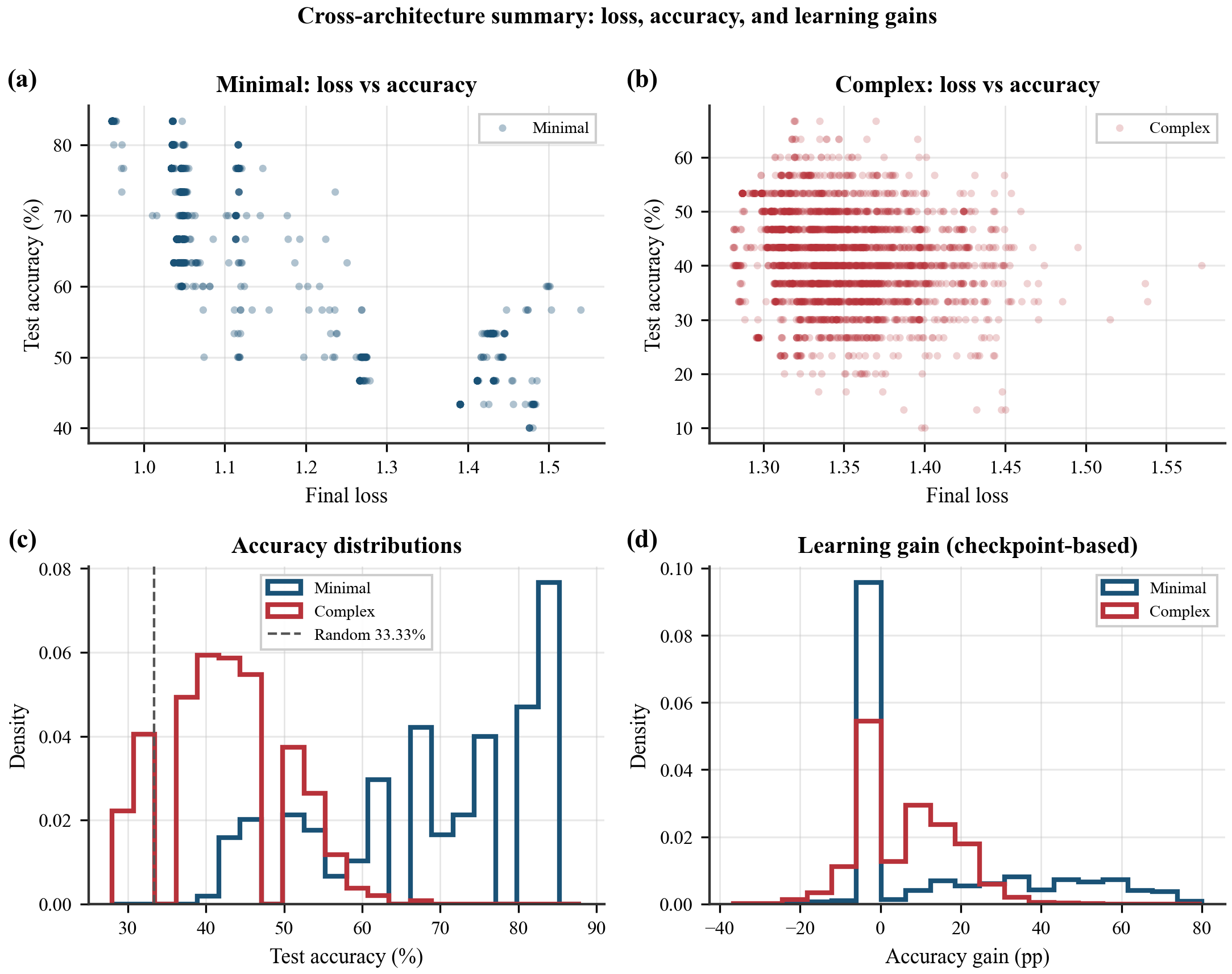}
 \caption{Statistical summary across minimal and complex experiments. Panels (c--d): normalized step histograms (not kernel density). Headline claims use peaks and Mann--Whitney $U$ (Section~\ref{sec:normality}).}
 \label{app:fig:statistical}
\end{figure*}

Figure~\ref{app:fig:statistical} presents multi-panel statistical analysis across both minimal and complex circuit experiments.

Summary:

\paragraph{Variance by architecture.}
Minimal circuits exhibit $\sigma \approx 9$--13\% (higher sensitivity); complex circuits $\sigma \approx 7$--10\% (moderate sensitivity). Fewer parameters (4 versus 16) amplify initialization effects on the minimal design.

\paragraph{Starting accuracy versus improvement.}
There is a weak negative correlation ($r = -0.23$): runs that start closer to good solutions show smaller measured improvement, which is common in gradient-free optimization with multiple local minima.

\paragraph{Significance versus baseline.}
In 77.5\% of runs, accuracy exceeds the random baseline significantly ($p < 0.001$), consistent with systematic learning rather than noise. Fifty seeds per configuration support stable comparisons.

\paragraph{Overfitting flags.}
Reported overfitting rates are 17.9\% on complex circuits (465 of 2,600 runs) and roughly 10\% on minimal circuits (estimated from the same diagnostics). These rates suggest reasonable generalization for small training sets ($n=120$).

\subsubsection{Seed Sensitivity Analysis}

The high seed count (50 per configuration) enables robust characterization of initialization sensitivity. Minimal circuits exhibit higher variance in part because there are only four base parameters versus sixteen on the complex design, the product feature map uses no CNOTs, and the loss landscape is shallow but sensitive to initialization.

Complex circuits show moderate variance despite more parameters, likely because eighteen CNOT gates create a more rugged landscape, the parity multiclass readout imposes a performance ceiling, and entanglement correlates parameters and reduces effective dimensionality.

\subsubsection{Performance Distribution Analysis}

The empirical distribution of improvements $\Delta$ is heavy-tailed, motivating quantile and peak reporting rather than Gaussian means. Roughly 60.7\% of runs show clear improvement ($\Delta>0$) with smooth loss decrease, while 39.3\% show limited gain, often from poor initializations in flat regions of the risk surface under the multiclass parity readout and a fixed iteration budget.

This pattern motivates reporting many random seeds, matching optimizers to architecture (L-BFGS-B on minimal designs, COBYLA on complex designs in our experiments), and using OQMD to improve outcomes across initializations.

\section{Classical baselines and outlook}
\label{app:discussion}

\subsection{Classical reference accuracies}
\label{app:classical_comparison}

Peak hold-out accuracies on Iris in this work span 56.67--96.67\% across the factorial and ladder configurations (Tables~\ref{tab:architecture_summary} and~\ref{tab:six_cnot_configs}): intermediate templates at six and nine CNOT reach 96.67\%, the minimal factorial peaks at 83.33\% with OQMD, and the 18-CNOT factorial peaks at 66.67\%. These values sit far below off-the-shelf classical models, which typically land in the mid-90\% range on the same dataset: for example, logistic regression, kernel SVMs, random forests, and small MLPs routinely exceed 95--98\% accuracy under comparable stratified splits.
The gap is expected: we fix Qiskit's default parity multiclass mapping, use only shallow circuits with at most a few dozen parameters, and optimize under a small-sample, noiseless simulator budget. The point of the comparison is not raw competitiveness but \textit{calibration}---readers should interpret our numbers as engineering trade-offs inside a constrained VQC design.

\subsection*{Outlook}
\label{app:future_work}

Future work should extend the same peak-focused reporting to larger datasets and calibrated noise models, revisit intermediate architectures with a full rotation factorial when resources allow, and, if finer learning curves are required, allocate additional compute for more frequent hold-out evaluation or alternative monitoring via training loss.

\end{appendices}

\section*{Declarations}

\paragraph{Funding.}
This work is supported by JSPS KAKENHI Grant Numbers JP24K14816 and ERATO ``Super Quantum Entanglement'' (Grant No.\ JPMJER2402) from JST.
This project also benefited from the exchange program between Binghamton University (SUNY) and the University of Electro-Communications (UEC), which supported M.\,A.\ Magid's residency at UEC during part of this work.

\paragraph{Competing interests.}
The authors have no competing interests to declare that are relevant to the content of this article.

\paragraph{Authors' contributions.}
M.\,A.\ Magid: methodology, software, formal analysis, investigation, visualization, writing---original draft, writing---review \& editing. M.\,Z.\,Ertem: writing---review \& editing. J.\ Suzuki: conceptualization, methodology, supervision, funding acquisition, writing---review \& editing. All authors approved the final manuscript.

\paragraph{Data availability.}
No new datasets were generated for this study. The Iris benchmark is publicly available through scikit-learn~\cite{pedregosa2011scikit}.

\paragraph{Code availability.}
The custom code used to train the variational circuits and produce the figures and tables is not yet deposited in a public repository. Before or upon acceptance, the authors intend to release a reproducibility package (scripts, configuration files, and summarized run outputs) in a version-controlled archive with a persistent identifier (e.g., Zenodo or an institutional GitHub repository). Until that release is public, code and materials sufficient to reproduce the reported results are available from the corresponding author on reasonable request.

\bibliographystyle{plainnat}

\end{document}